\documentclass[prd,twocolumn,a4paper,superscriptaddress]{revtex4}
\usepackage{graphicx}

\begin{document}

\newcommand {\R}{{\mathcal R}}
\newcommand{\al}{\alpha}

\title{Measuring $\alpha$ in the Early Universe:\\
CMB Temperature, Large-Scale Structure and Fisher Matrix Analysis}

\author{C.J.A.P. Martins}
\email[Electronic address: ]{C.J.A.P.Martins@damtp.cam.ac.uk}
\affiliation{Centro de Astrof\'{\i}sica da Universidade do Porto, R. das
Estrelas s/n, 4150-762 Porto, Portugal}
\affiliation{Department of Applied Mathematics and Theoretical Physics,
Centre for Mathematical Sciences,\\ University of Cambridge,
Wilberforce Road, Cambridge CB3 0WA, United Kingdom}
\affiliation{Institut d'Astrophysique de Paris, 98 bis Boulevard Arago,
75014 Paris, France}
\author{A. Melchiorri}
\email[Electronic address: ]{melch@astro.ox.ac.uk}
\affiliation{Department of Physics, Nuclear \& Astrophysics Laboratory,
University of Oxford, Keble Road, Oxford OX1 3RH, United Kingdom}
\author{R. Trotta}
\email[Electronic address: ]{trotta@amorgos.unige.ch}
\affiliation{D\'epartement de Physique Th\'eorique, Universit\'e de
Gen\`eve, 24 quai Ernest Ansermet, CH-1211 Gen\`eve 4, Switzerland}
\author{R. Bean}
\email[Electronic address: ]{r.bean@ic.ac.uk}
\affiliation{Theoretical Physics, The Blackett Laboratory, Imperial
College, Prince Consort Road, London SW7 2BZ, United Kingdom}
\author{G. Rocha}
\email[Electronic address: ]{graca@mrao.cam.ac.uk}
\affiliation{Astrophysics Group, Cavendish Laboratory,
Madingley Road, Cambridge CB3 0HE, United Kingdom}
\affiliation{Centro de Astrof\'{\i}sica da Universidade do Porto, R. das
Estrelas s/n, 4150-762 Porto, Portugal}
\author{P. P. Avelino}
\email[Electronic address: ]{pedro@astro.up.pt}
\affiliation{Centro de Astrof\'{\i}sica da Universidade do Porto, R. das
Estrelas s/n, 4150-762 Porto, Portugal}
\affiliation{Departamento de F\'{\i}sica da Faculdade de Ci\^encias
da Universidade do Porto, R. do Campo Alegre 687, 4169-007 Porto, Portugal}
\author{P.T.P. Viana}
\email[Electronic address: ]{viana@astro.up.pt}
\affiliation{Centro de Astrof\'{\i}sica da Universidade do Porto, R. das
Estrelas s/n, 4150-762 Porto, Portugal}
\affiliation{Departamento de Matem\'atica Aplicada da Faculdade de Ci\^encias
da Universidade do Porto, Rua do Campo Alegre 687, 4169-007 Porto, Portugal}

\date{8 March 2002}

\begin{abstract}
We extend our recent work on the effects of a time-varying fine-structure
constant $\alpha$ in the cosmic microwave background, by providing a
thorough analysis of the degeneracies between $\alpha$ and the other
cosmological parameters, and discussing ways to break these with both
existing and/or forthcoming data.
In particular, we present the state-of-the-art CMB constraints on $\alpha$,
through a combined analysis of the BOOMERanG, MAXIMA and DASI datasets.
We also present a novel discussion of the constraints on $\alpha$ coming
from large-scale structure observations, focusing in particular on
the power spectrum from the 2dF survey.
Our results are consistent with no variation in $\alpha$ from the
epoch of recombination to the present day, and restrict any such
variation to be less than about $4\%$.
We show that the forthcoming MAP and Planck experiments will
be able to break most of the currently existing degeneracies between $\alpha$
and other parameters, and measure $\alpha$ to better than percent accuracy.
\end{abstract}
\preprint{DAMTP-2002-14}
\maketitle

\section{\label{intr}Introduction}
The search for observational evidence for time or space variations of
the `fundamental' constants that can be measured in our
four-dimensional world is an extremely exciting area of
current research, with several independent claims of detections in different
contexts emerging in the past year or so, together with other
improved constraints
\cite{Murphy:2000pz,Webb:2000mn,Murphy:2000ns,Murphy:2001nu,Avelino:2001nr,
Fujii:2002bi,Ivanchik:2001ji}. We will review these in Sect. \ref{obs}.

Most of the current efforts have been concentrating on the
fine-structure constant, $\alpha$, both due to its obviously
fundamental role and due to the availability of a series of
independent methods of measurement. Noteworthy among these is
the Cosmic Microwave Background
(CMB) \cite{Avelino:2000oo,Avelino:2000ea,Battye:2000ds,Avelino:2001nr}.
The latest available CMB results \cite{Avelino:2001nr}
yield a one-sigma indication of a smaller $\alpha$ in the past,
but are consistent with no variation at the two-sigma level.
However, these results are somewhat weakened by the existence of
various important degeneracies in the data, and furthermore not
everybody agrees on which and how strong these degeneracies
are \cite{Avelino:2000ea,Battye:2000ds,Huey:2001ku}.

Here we aim to clarify this issue by analyzing these
possible degeneracies in some detail, mainly by means of a
Fisher Matrix Analysis (FMA), see Sect. \ref{fma}.
We will emphasize that there
are crucial differences between `theoretical' degeneracies
(due to simple physical mechanisms) and `experimental'
degeneracies (due to the fact that each CMB experiment
only probes a limited range of scales, and that the experimental
errors are scale-dependent). We will also show how such
degeneracies can be eliminated either by using complementary
data sets (such as large-scale structure constraints, see
Sect. \ref{sig8}) or by acquiring better data (such as that to
be obtained by MAP and Planck). We present our
conclusions in Sect. \ref{conc}.

In a companion paper \cite{Rocha:2002} we will discuss a further
way in which these degeneracies can be broken, namely by including
information from CMB polarization.

\section{\label{obs}The Present Observational Status}

The recent explosion of interest in the study of varying constants is
mostly due to the results of Webb and
collaborators \cite{Murphy:2000pz,Webb:2000mn,Murphy:2000ns,Murphy:2001nu}
of a $4\sigma$ detection of a fine-structure constant that was smaller
in the past,
\begin{equation}
\frac{\Delta\alpha}{\alpha} = (-0.72 \pm 0.18)\times 10^{-5}\,,\quad
z\sim0.5-3.5\,; \label{webb}
\end{equation}
indeed, more recent work \cite{Webb:2001} provides an even stronger
detection. These results are obtained through comparisons of various
transitions (involving various different atoms)
in the laboratory and in quasar absorption systems, using the fact that
the size of the relativistic corrections
goes as $(\alpha Z)^{2}$. A number of tests for possible systematic
effects have been carried out, all of which have been found either
not to affect the results or to make the detection even stronger if
corrected for.

A somewhat analogous (though simpler) technique uses molecular hydrogen
transitions in damped Lyman-$\alpha$ systems to measure the ratio of
the proton and electron masses, $\mu=m_p/m_e$ (using the fact that
electron vibro-rotational lines depend on the reduced mass of the
molecule, and this dependence is different for different transitions).
The latest results \cite{Ivanchik:2001ji} using two systems at
redshifts $z\sim2.3$ and $z\sim3.0$ are
\begin{equation}
\frac{\Delta\mu}{\mu} = (5.7 \pm 3.8)\times 10^{-5}\ 
\label{pett1}
\end{equation}
or
\begin{equation}
\frac{\Delta\mu}{\mu} = (12.5 \pm 4.5)\times 10^{-5}\,,
\label{pett2}
\end{equation}
depending on which of the (two) available tables of `standard'
laboratory wavelengths is used. This implies a $1.5 \sigma$
detection in the more conservative case, though it also casts some
doubts on the accuracy of the laboratory results, and on the
influence of systematic effects in general.

We should also mention a recent re-analysis \cite{Fujii:2002bi}
of the well-known Oklo bound \cite{Damour:1996zw}. Using new
Samarium samples
collected deeper underground (aiming to minimize contamination), these
authors again provide two possible results for both $\alpha$ and the
analogous coupling for the strong nuclear force, $\alpha_s$,
\begin{equation}
\frac{\dot\alpha}{\alpha} \sim \frac{\dot\alpha_s}{\alpha_s}=
(0.4 \pm 0.5)\times 10^{-17} yr^{-1}
\label{oklo1}
\end{equation}
or
\begin{equation}
\frac{\dot\alpha}{\alpha} \sim \frac{\dot\alpha_s}{\alpha_s}=
-(4.4 \pm 0.4)\times 10^{-17} yr^{-1}\,.
\label{oklo2}
\end{equation}
Note that these are given as rates of variation, and effectively
probe timescales corresponding to a cosmological redshift of about
$z\sim 0.1$. Unlike the case above, these two values correspond
to two possible physical branches of the solution. See
\cite{Fujii:2002bi} for a discussion of why this method yields
two solutions (and also note that these results have opposite signs
relative to previously published ones \cite{Fujii:1998kn}).
While the first of these branches provides a null result,
(\ref{oklo2}) is a strong detection of an $\alpha$ that was
{\it larger} at $z\sim0.1$, that is a relative variation
that is opposite to Webb's result (\ref{webb}). Even though there
are some hints (coming from the analysis of other Gadolinium
samples)
that the first branch is preferred, this is by no means settled
and further analysis is required to verify it.

Still we can speculate about the possibility that the second branch
turns out to be the correct one. Indeed this would definitely
be the most exciting possibility. While in itself this wouldn't
contradict Webb's results (since Oklo probes much smaller
redshift and the suggested magnitude of the variation is smaller
than that suggested by the quasar data), it would have
striking effects on the theoretical modelling of such variations.
In fact, proof that $\alpha$ was once larger than today's value
would sound the death knell for any theory which models
the varying $\alpha$ through a scalar field whose behaviour
is akin to that of a dilaton. Examples include Bekenstein's theory
\cite{Bekenstein:1982eu} or simple variations thereof
\cite{Sandvik:2001rv,Olive:2001vz}.
Indeed, one can quite easily see
\cite{Damour:1993id,Santiago:1998ae} that in any such model
having sensible cosmological parameters and obeying other
standard constraints 
$\alpha$ must be a monotonically increasing function of time.
Since these dilatonic-type models are arguably the simplest
and best-motivated models for varying alpha from a particle
physics point of view, any evidence against them would be
extremely exciting, since it would point towards the presence
of significantly different, yet undiscovered physical mechanisms.

Finally, we also mention that there have been recent proposals
\cite{Braxmaier:2001ph} of more accurate laboratory tests of
the time independence of $\alpha$ and the ratio of the
proton and electron masses $\mu$ using monolithic
resonators, which could improve current bounds by an order of
magnitude or more.

However, given that there are both theoretical and experimental
reasons to expect that any recent variations will be small, it
is important to develop tools allowing us to measure $\alpha$ in
the early universe, as variations with respect to the present value
could be much larger then.

In what follows
we focus on the analysis of CMB data allowing for possible
variations of the fine-structure constant. In our previous
work \cite{Avelino:2001nr}, we have carried out a joint
analysis using the most recent CMB (BOOMERanG and DASI) and
big-bang nucleosynthesis (BBN) data, finding evidence at the one sigma
level for a smaller alpha in the past (at the level of $10^{-2}$ or
$10^{-3}$), though at the two sigma level the results were
consistent with no variation. However, as can be seen by comparing
with earlier work \cite{Avelino:2000ea,Battye:2000ds} (and has
also been discussed explicitly in these papers), these
results are quite strongly dependent on both the observational
datasets and the priors one uses.

Regarding this latter issue, we point out that a recent \cite{Coc:2002tr}
improved analysis of standard BBN (focusing mostly on
nuclear physics aspects) suggests that ${}^7Li$ could lead to more stringent
constraints on the baryonic density of the universe ($\Omega_b$) than deuterium.
The point made by the authors
is that ${}^7Li$ is effectively a better baryometer than $D$,
because of difficulties in obtaining (extrapolated) primordial
abundances of the latter. They 
then obtain values for $w_{b}\equiv\Omega_{b}h^2$
that are considerably lower than the standard
ones. These results are also corroborated by \cite{Cyburt:2001pp}.
Using these results as a prior would transform our previous
result \cite{Avelino:2001nr} into a detection of a varying $\alpha$
at more than two sigma.

In any case, previous analyses of CMB data allowing for a varying
$\alpha$ \cite{Avelino:2000ea,Battye:2000ds,Avelino:2001nr} have
revealed some interesting degeneracies between $\alpha$ and other
cosmological parameters, such as $\Omega_b$ or $H_0$. On the other
hand, a recent `brute-force' exploration of a particular sector
of parameter space (including quintessence models) \cite{Huey:2001ku} seems to
claim results on degeneracies between the various
parameters \cite{Avelino:2000ea,Battye:2000ds,Avelino:2001nr}.

While the two approaches are not really comparable (\cite{Huey:2001ku}
being rather more simplistic, as it uses no actual data and
has somewhat unclear criteria for the presence of
a degeneracy), this discrepancy begs the question of whether
the degeneracies found in \cite{Avelino:2000ea,Battye:2000ds,Avelino:2001nr}
are real `physical' and fundamental degeneracies, which will remain at
some level, no matter how much more accurate data one can get, or
if they are simply degeneracies in the data, which won't
necessarily be there in other (better) datasets. And a related
question is, of course, assuming that the degeneracies are
significant, how can one get around them. We will address these issues
in the following sections.

\section{\label{sig8}Current CMB and Large-Scale Structure Constraints}

Here we present an up-to-date analysis of the Cosmic Microwave Background
constraints on varying $\alpha$ as well as, for the first time,
an analysis of its
effects on the large-scale structure (LSS) power spectrum.

Even though this may not be entirely obvious, a varying $\alpha$
will have an effect on the matter power spectrum. The simplest
way to understand this is to interpret the variation in
$\alpha$ as being due to a variation in the speed of light $c$
(which one is always free to do \cite{Avelino:1999is}).

A variation in $\alpha$ affects the matter power spectrum to the extent 
that it changes the horizon size, hence the turnover scale in the matter
power spectrum.
Allowing for a variation in $\alpha$, this is not only a 
function of $\Omega_{m}$, $\Omega_{B}$ and $h$ but of $\alpha$ as 
well, through the dependence of the recombination
epoch on $\alpha$. Therefore varying alpha will produce 
a change in the turnover point position $k_{rec}$ of the matter 
power spectrum, hence a shift of the curve sideways, and therefore a 
change on the value of $\sigma_{8}$. For example a decrease in $\alpha$ 
shifts this turnover scale to smaller $k$, hence allowing for a decrease 
in $\sigma_8$.

By plotting the transfer functions (generated with a modified
\cite{Avelino:2000ea} version of the {\sc CMBFAST} \cite{sz}
code which includes the 
effects of a varying $\alpha$) we find that this effect is fairly small.
For $\Omega_{m}=0.3$ and $\Omega_B=0.05$ and keeping all other cosmological
parameters fixed a variation of $\alpha$ by say $10\%$ from its standard 
value produces variations in the transfer function which are at most $5\%$ 
in restricted regions of $k$ (the effect on the value of $\sigma_8$ is 
even smaller).

On the other hand a change in $\alpha$ will modify the
\textit{height} of the first 
peak of the CMB power spectrum through the (early) ISW effect. 
This effect also depends on $\Omega_m h^{2}$. This 
illustrates the interplay between a varying $\alpha$,  $\Omega_{m}$ and 
the value of $\sigma_{8}$.
Further effects of a varying $\alpha$ in the CMB are
a slight change in the \textit{position} of the first peak due to
the aforementioned change in the horizon size, plus a variation in
the high-$\ell$ damping (due to the finite thickness of the last-scattering
surface) which are also dependent on a number of cosmological parameters 
other than $\alpha$.

It should be emphasized that although these CMB and LSS constraints are
in some sense complementary, and can help break degeneracies
by determining other cosmological parameters, they certainly
can not be blindly combined together, since the range of
cosmological epochs (or redshifts) to which they are sensitive
is somewhat different.

\subsection{CMB data analysis}

We compare  the recent CMB observations with a set of 
flat models with parameters sampled as follows (the value in brackets
is the step size): 
\begin{equation}
\Omega_m=0.1\ldots (0.1)\ldots 1.0
\label{params1}
\end{equation}
\begin{equation}
\Omega_b=\Omega_{m}-\Omega_{cdm}=0.009\ldots (0.003)\ldots 0.036
\label{params2}
\end{equation}
\begin{equation}
\frac{\Delta \alpha}{\alpha}=0.80\ldots (0.01)\ldots 1.10
\label{params3}
\end{equation}
\begin{equation}
h=0.40\ldots (0.05)\ldots 0.90
\label{params4}
\end{equation}
\begin{equation}
n_s=0.70\ldots (0.05)\ldots 1.30\, .
\label{params5}
\end{equation}

We rescale the amplitude of fluctuations by a
pre-factor $C_{10}$, in units of $C_{10}^{COBE}$, with 
$0.50 < C_{10} < 1.40$. Finally, we assumed a negligible re-ionization
and an optical depth $\tau_c\sim0$. 
This is in agreement with recent estimates on the redshift of re-ionization
$z_{re}\sim 6 \pm 1$ (see e.g. \cite{gnedin}).

The theoretical models are computed using a modified version
of the publicly available {\sc CMBFAST} program~\cite{sz}, accounting
for the effects of a varying $\alpha$, 
and are compared with the recent 
BOOMERanG-98, DASI and MAXIMA-1 results.
The power spectra from these experiments were estimated in 
$19$, $9$ and $13$ bins respectively, spanning the range
$25 \le \ell \le 1150$.

For the DASI and MAXIMA-I experiment we use the publicly available
correlation matrices and window functions.
For the BOOMERanG experiment we assign a constant value
for the spectrum in each bin $\ell(\ell+1)C_{\ell}/2\pi=C_B$, 
we approximate the signal $C_B$ inside
the bin to be a Gaussian variable and we consider $\sim 10 \%$ 
correlations between contiguous bins.
The likelihood for a given cosmological model is then
 defined by 

\begin{equation}
-2{\rm ln} {\cal L}=(C_B^{th}-C_B^{ex})M_{BB'}(C_{B'}^{th}-C_{B'}^{ex})\, ,
\label{likely01}
\end{equation}
where $M_{BB'}$ is the Gaussian curvature of the likelihood 
matrix at the peak. 
We consider $10 \%$, $4 \%$  and $5 \%$ Gaussian distributed 
calibration errors for the BOOMERanG-98, DASI 
and MAXIMA-1 experiments respectively and 
we included the beam uncertainties by the analytical marginalization
method presented in \cite{sara}.
We also include the COBE data using Lloyd Knox's RADPack packages.

\subsection{LSS data analysis}

In what follows, we will add to the CMB data the real-space 
power spectrum of galaxies in the 2dF 100k galaxy redshift survey using the
data and window functions of the analysis of \cite{thx}.

To compute ${\cal L}^{2dF}$ we, evaluate $p_i = P(k_i)$, 
where $P(k)$ is the theoretical matter power spectrum 
and $k_i$ are the $49$ k-values of the measurements in \cite{thx}. 
Therefore we have

\begin{equation}
-2ln{\cal L}^{2dF} = \sum_i [P_i-(Wp)_i]^2/dP_i^2\, ,
\label{likely02}
\end{equation}
where $P_i$ and $dP_i$ are the measurements and error bars in \cite{thx}
and $W$ is the reported $27 \times 49$ window matrix.
We restricted the analysis to a range of scales where the fluctuations
are assumed to be in the linear
regime ($k < 0.02 h^{-1}Mpc$).
When combining with the CMB data, we marginalize over a bias $b$ 
considered as additional free parameter.

We will also include information on $\sigma_8$, the RMS mass fluctuation
in spheres of $8 h^{-1}Mpc$, obtained from local cluster number counts.
There is presently no consensus on the correct value of this observable,
mainly because of systematics in the calibration between
cluster virial mass and temperature. For convenience of
analysis, we consider $2$ values: an {\it high} value 
$\sim \Omega_m^{0.6} \sigma_8=0.50 \pm 0.05$ in agreement with the results 
of \cite{pierpa,eke} and a {\it lower} one, 
$\sim \Omega_m^{0.6} \sigma_8=0.40 \pm 0.05$ 
following the analysis of \cite{s8eljak,liddle}.

We attribute a likelihood 
to each value of $\delta \alpha / \alpha$ by marginalizing over 
the {\it nuisance} parameters. 
We then define our $68\%$ ($95 \%$), 
confidence levels to be where the integral of the 
likelihood is $0.16$ ($0.025$) and $0.84$ 
($0.975$) of the total value (see e.g. \cite{melchiorri}).

\begin{figure}
\includegraphics[width=3in,keepaspectratio]{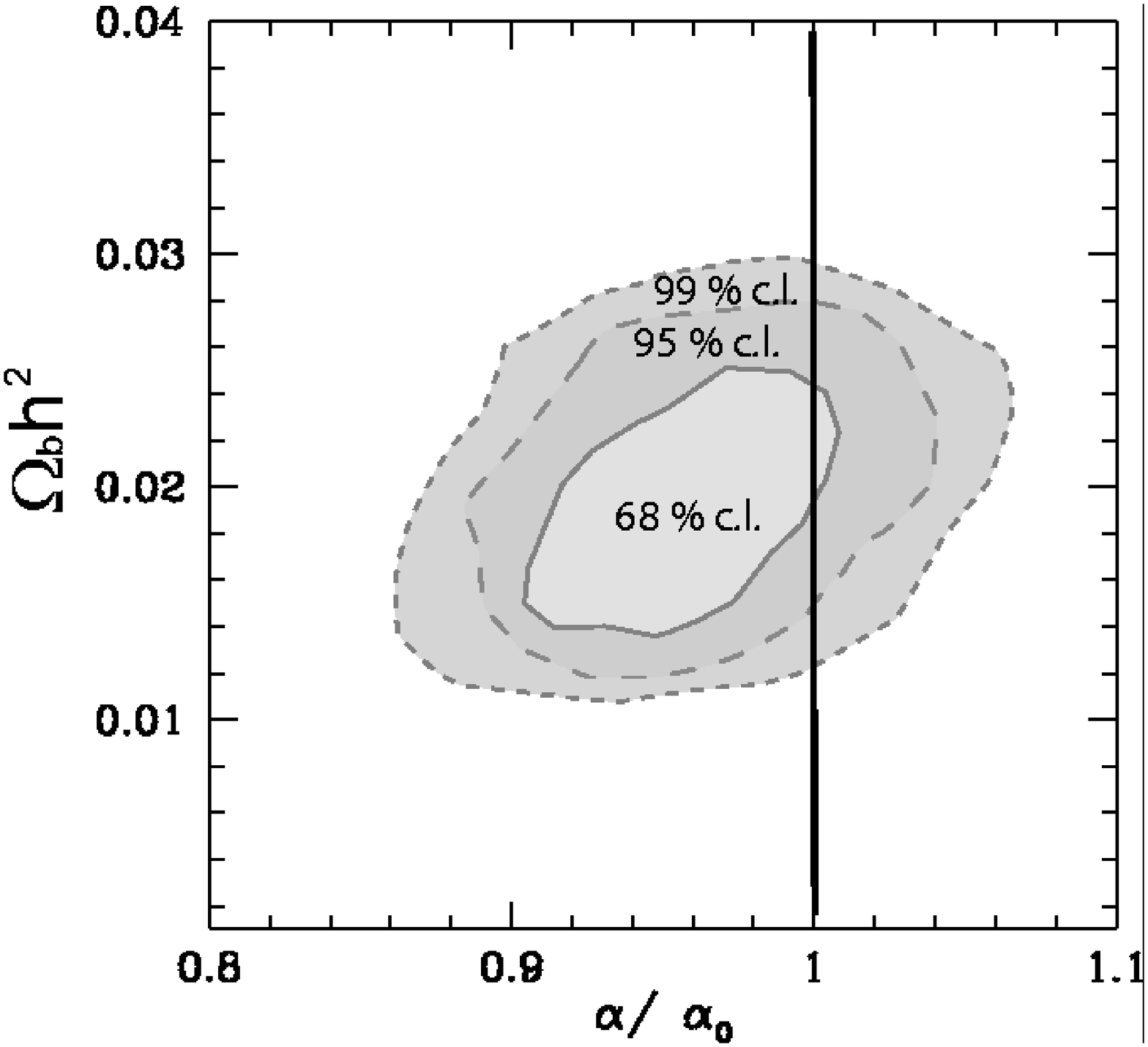}
\includegraphics[width=3in,keepaspectratio]{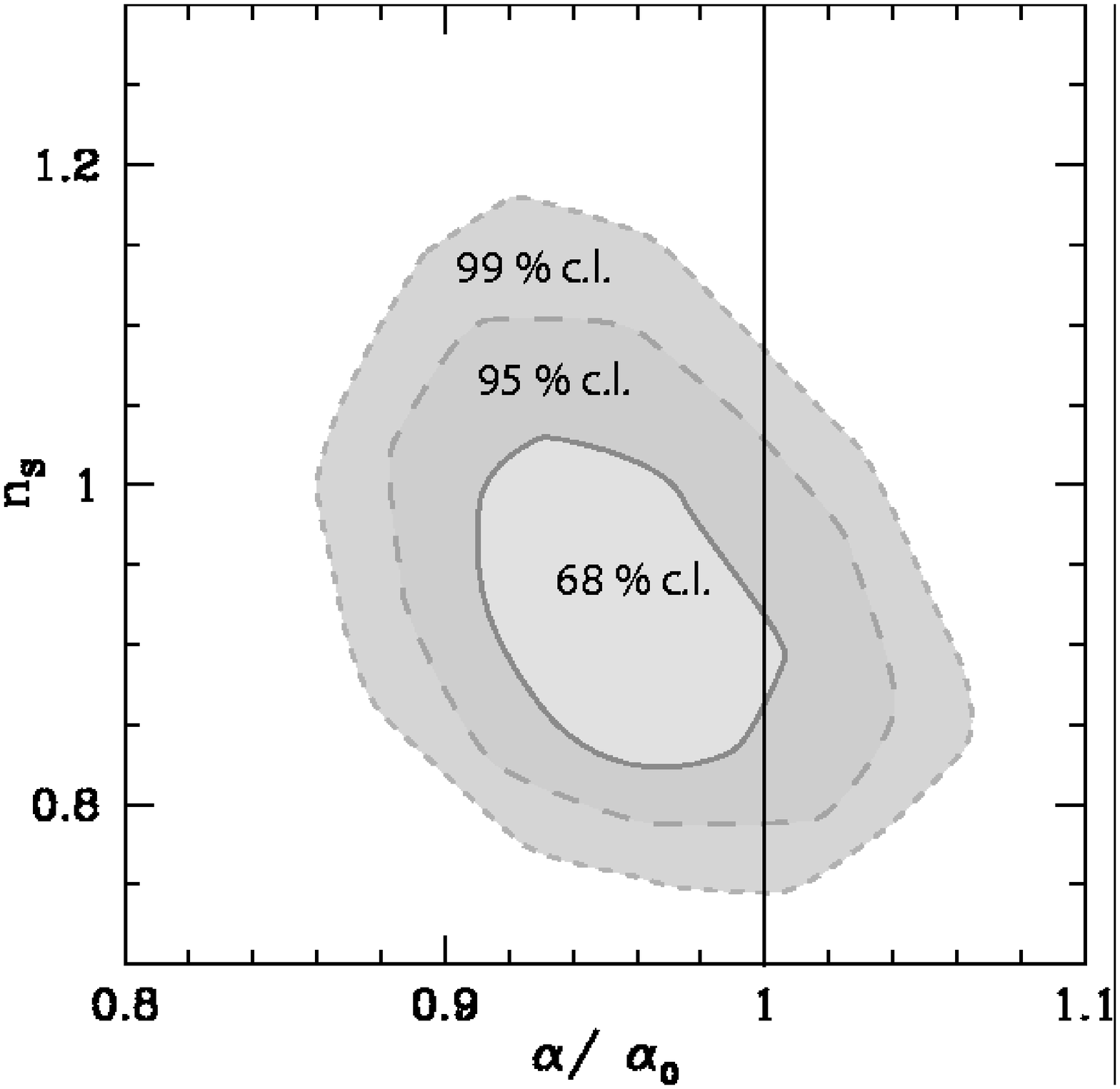}
\includegraphics[width=3in,keepaspectratio]{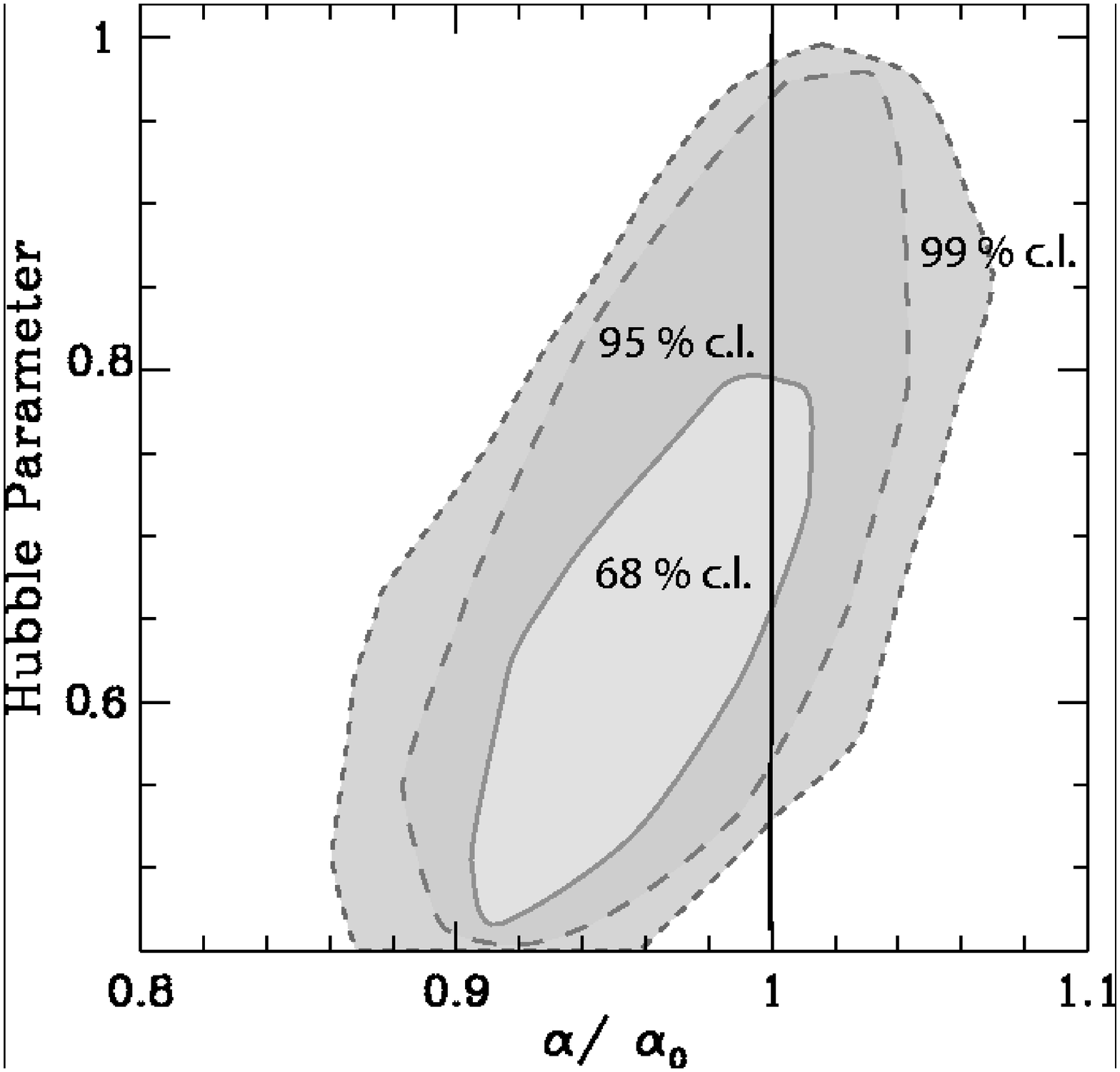}
\caption{\label{fig1}The current combined
CMB constraints on the $\omega_b-\alpha / \alpha_0$,
$n_s-\alpha / \alpha_0$ and $h-\alpha / \alpha_0$ planes.}
\end{figure}

\begin{table}
\caption{\label{alphas}Current cosmological constraints on the
variation of the fine-structure constant (marginalizing over other
parameters) for various different priors.}
\begin{ruledtabular}
\begin{tabular}{|l|c|}
\hline
\hline
Prior &$\alpha /\alpha_0$ (95 \% c.l.)\\ \hline
$h=0.65\pm0.2$ &$0.95_{-0.06}^{+0.07}$ \\ \hline
BBN $\omega_b=0.02\pm0.002$ &$0.96_{-0.05}^{+0.06}$ \\ \hline
HST $h=0.71\pm0.08$ &$0.98_{-0.05}^{+0.05}$ \\ \hline
SN-Ia &$0.95_{-0.06}^{+0.07}$ \\ \hline
$\Omega_m^{0.6}\sigma_8=0.50\pm0.05$ &$1.01_{-0.04}^{+0.04}$ \\ \hline
$\Omega_m^{0.6}\sigma_8=0.40\pm0.05$ &$0.99_{-0.05}^{+0.05}$ \\ \hline
2dF &$0.98_{-0.05}^{+0.04}$ \\ \hline
\end{tabular}
\end{ruledtabular}
\end{table}

\subsection{Results}

In a previous work \cite{Avelino:2001nr} we produced 
likelihood contours in the $\Omega_b h^2- \alpha / \alpha_0$ plane
by analyzing the recent BOOMERanG and DASI CMB datasets and
by including two priors to the analysis: flatness and 
$h=0.65 \pm 0.2$. Our results were
consistent with the baryon abundance obtained from
big-bang nucleosynthesis (BBN) and we constrained variations 
in $\alpha$ at $z \sim 1000$ at level of about $10 \%$.
In Fig. \ref{fig1} we plot constraints on this plane, as well as on the 
$n_s - \alpha / \alpha_0$ and $h - \alpha / \alpha_0$ planes
with a similar analysis, but including this time the MAXIMA-I 
dataset.

\begin{figure}
\includegraphics[width=3in,keepaspectratio]{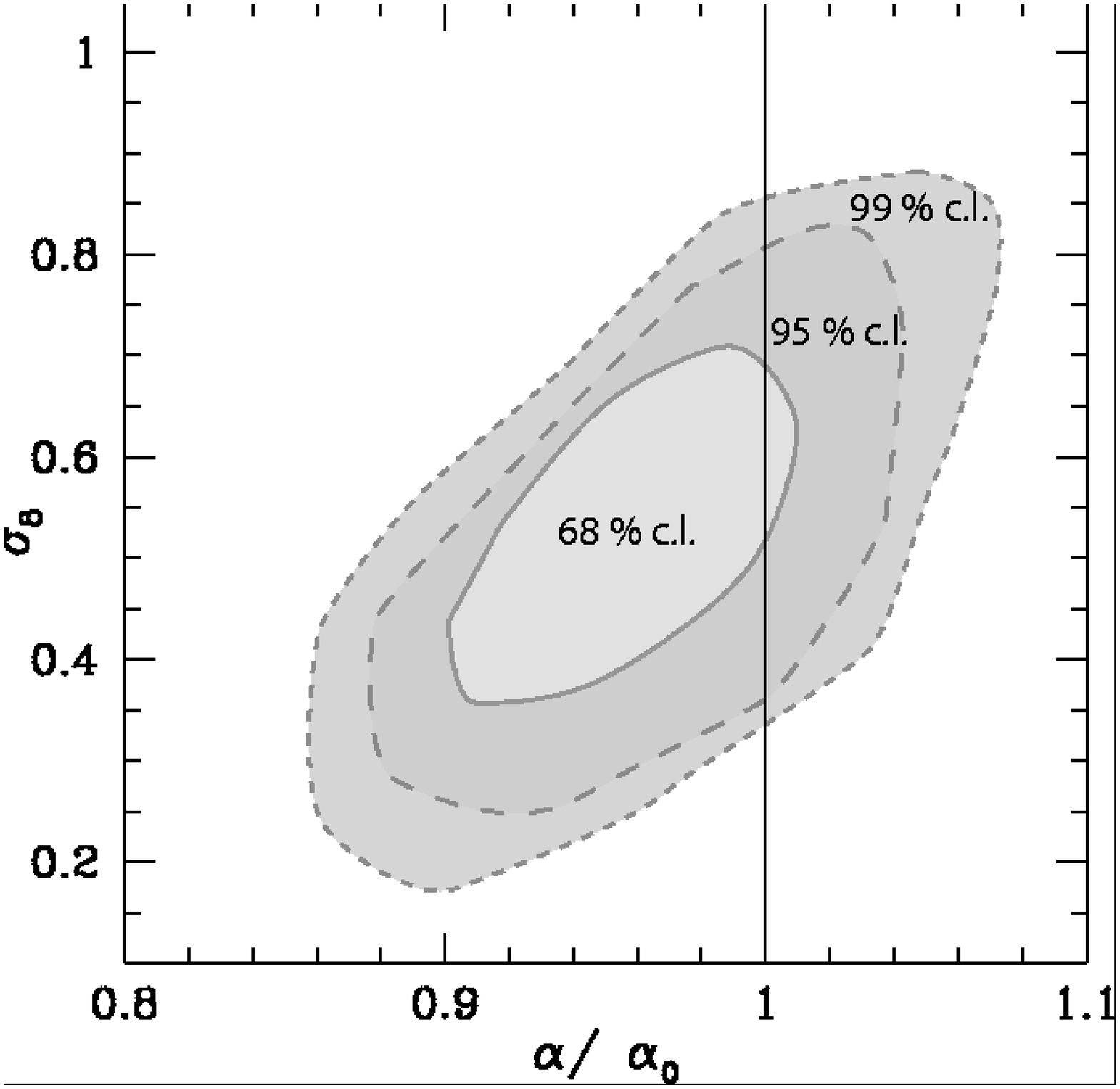}
\includegraphics[width=3in,keepaspectratio]{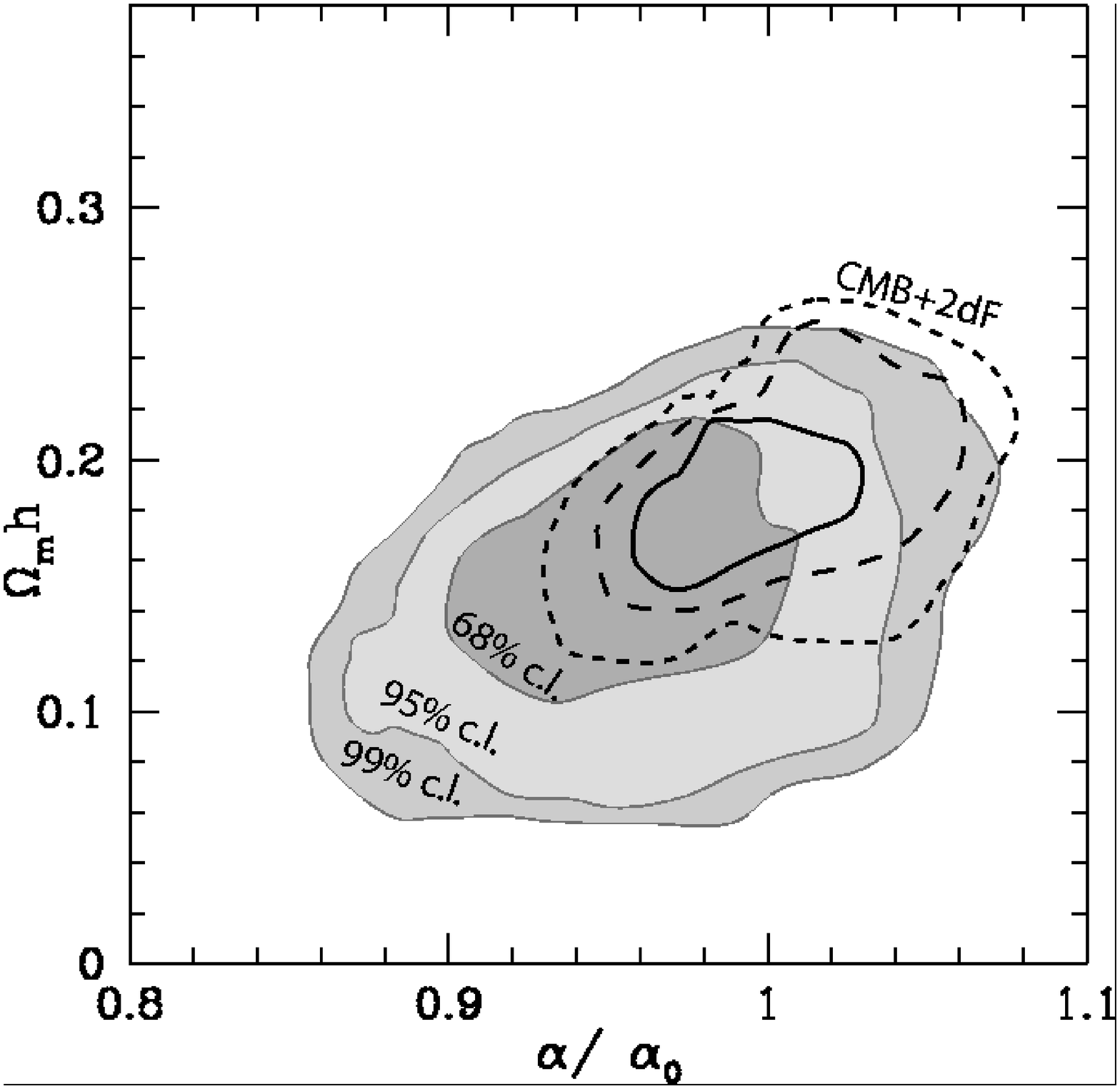}
\caption{\label{fig2}The current combined
CMB constraints on the $\sigma_8-\alpha / \alpha_0$
and $\Omega_m h -\alpha / \alpha_0$ planes. The latter panel also
shows the results of combining the CMB and 2dF datasets.}
\end{figure}

From these results we can see that
the inclusion of the MAXIMA-I data doesn't significantly change our
previous constraints. In fact, even if the analysis of the MAXIMA-I
data alone suggests a {\it higher} value of the baryon fraction
($\Omega_bh^2 \sim 0.030 \pm 0.005$ \cite{stompor}), the combined
analysis with DASI and BOOMERanG still suggests a {\it low} value
of $\Omega_bh^2 \sim 0.023$. This result is in agreement
with previous analysis, e.g. \cite{thx}.

From last two panels of Fig. \ref{fig1} we also see that, in the
set of models we are considering, there is a clear correlation 
between variations in $\alpha$ and changes in the scalar 
spectral index $n_s$ and the Hubble parameter $h$.
We will see in the next section that this degeneracy can
be broken by the future and more accurate measurements
from satellite experiments like MAP \cite{MAP} or the
Planck Surveyor \cite{Planck}.

However, since variations in $n_s$ and $h$ affect the shape, position
and amplitude of the matter power spectrum (irrespective of changes in 
$\alpha$), we can in principle use the data from galaxy clustering and 
local cluster abundances in order to break these degeneracies 
and infer stronger constraints on $\alpha$.
Regarding the RMS amplitude of mass fluctuations on scales of
$8 h^{-1}Mpc$, the two effects are competitive: an increase in 
$n_S$ will increase $\sigma_8$, while lowering $h$
decreases it. As we can see from the first panel of Fig. \ref{fig2},
the effect from $h$ is stronger and 
the final result is that a decrease in $\alpha$ 
generally allows for lower amplitudes of $\sigma_8$.

Furthermore, as we can see from Fig. \ref{fig2}, a certain degree of
degeneracy is present in the CMB data between $\alpha$ and the shape
parameter $\Omega_m h$. This degeneracy can be optimally
broken by incorporating the data from the 2dF galaxy survey.
As we can see in the bottom panel, including the 2dF data shrinks the
contours around $\alpha /\alpha_0 \sim 1$ and $\Omega_mh \sim 0.2$,
In other words, there is a clear distinction between the CMB and LSS
data: while for the CMB data a negative variation of $\alpha$ is
preferred, the opposite happens for the LSS data. When the two
datasets are combined, the best-fit model, in the $\Omega_m h-\alpha/\alpha_0$
plane, is quite close to the standard one.

We should emphasize at this point
that in doing this we are not combining direct constraints
on the parameter $\alpha$ itself, obtained through both methods, to
obtain a tighter constraint. As mentioned above this \textit{can not be done},
since the CMB and LSS analyses are sensitive to the values of $\alpha$
at different redshift ranges, so there is no reason why these values
should be the same. Additionally, there is no well-motivated theory that
could relate such variations at different cosmological epochs. All that
one could do at this stage would be to assume some toy model where a
certain behaviour would occur, but this would mean introducing various
additional parameters, thus weakening the analysis. Hence we chose not
to pursue this path and leave the analysis as model-independent as possible.

What we are doing is using additional information (which is also
sensitive to $\alpha$) to better constrain other parameters in the
underlying cosmological model, such as $n_s$, $h$ and the densities of
various matter components, which we \textit{can} reliably assume are
unchanged throughout the cosmological epochs in question. In other words,
we are simply selecting more stringent priors for our analysis, in a
self-consistent way.

The constraints obtained by combined analysis are reported
in Table \ref{alphas}.
Our main result from this table is that, as one would expect,
when constraints from 
other and independent cosmological datasets are included in the
CMB analysis, the constraints on variations on $\alpha$ become
significantly stronger.

\section{\label{fma}Fisher Matrix Analysis}

The precision with which the forthcoming satellite 
experiments MAP \cite{MAP} and Planck \cite{Planck} will be able to
determine variations in $\al$ can be readily 
estimated with a Fisher Matrix Analysis (FMA).
Some authors have already performed such an analysis in the past 
\cite{Hannestad:99,Kaplinghat:99}: however, their analysis was based on a 
different set of cosmological parameters and assumed cosmic variance 
limited measurements. In our FMA we also take into account 
the expected performance of the MAP and Planck satellites and we make use of a 
cosmological parameters set which is well adapted for limiting numerical 
inaccuracies. Furthermore, the FMA can provide useful insight into 
the degeneracies among different parameters, with minimal computational effort.

\subsection{Analysis Setup}

We characterize the cosmological
model by a 
7 dimensional parameter set, given by
\begin{equation}
{\bf \Theta} = (\al, \omega_b, \omega_m, \omega_\Lambda, \R, n_s, Q), 
\end{equation} 
where $\omega_b \equiv \Omega_b h^2$ is the physical baryonic density, 
$\omega_m \equiv (\Omega_{\rm cdm} +\Omega_b)h^2$ the energy
density in matter and $\omega_\Lambda$ the energy density
due to a cosmological constant. 
Here $h$ denotes the Hubble parameter today, $H_0 \equiv 100h$ 
km s$^{-1}$ Mpc$^{-1}$. The quantity 
$\R \equiv \ell_{\rm ref}  / \ell$ 
is the `shift' parameter (see \cite{melch:01,Bowen:01}
and references therein), which gives the position of the 
acoustic peaks with respect to a flat, $\Omega_\Lambda = 0$ reference model,
$n_s$ is the scalar spectral index and \mbox{$Q
= < \ell (\ell + 1) C_\ell > ^{1/2}$} denotes the overall
normalization, where the mean is taken over the multipole range
$2 \leq \ell \leq 2000$.

The shift parameter $\R$ depends on $\Omega_m$, on the curvature
$\Omega_{\kappa} \equiv 1 - \Omega_{\Lambda} - \Omega_m - \Omega_{\rm rad}$
through
\begin{eqnarray}
\label{eq:def_r}
\R &=& 2 \left( 1 - \frac{1}{\sqrt{1 + z_{\rm dec} }} \right) \nonumber \\
&& \times \frac{\sqrt{| \Omega_{\kappa}| }}{ \Omega_m}
\frac{1}{\chi(y)}
\left[ \sqrt{\Omega_{\rm rad} +
\frac{\Omega_m}{1 + z_{\rm dec} } } - \sqrt{\Omega_{\rm rad}} \right] ,
\end{eqnarray}
where $z_{\rm dec} $ is the redshift of decoupling, $\Omega_{\rm rad}$ is
the energy parameter due to radiation
($\Omega_{\rm rad}=4.13 \cdot 10^{-5}/h^2$ for photons and 3 neutrinos) and
\begin{eqnarray}
\label{eq:ydef}
y &=& \sqrt{|\Omega_{\kappa}|}\int_0^{z_{\rm dec} } \, dz\\
&& {[\Omega_{\rm rad} (1+z)^4 +
\Omega_m(1+z)^3+\Omega_{\kappa}(1+z)^2+\Omega_{\Lambda}]^{-1/2}}. \nonumber
\end{eqnarray}
The function $\chi(y)$ depends on the curvature of the universe and is
$y$, $\sin(y)$ or $\sinh(y)$ for flat, closed or open models,
respectively.
Inclusion of the shift parameter $\R$ into our set of parameters takes into
account the geometrical degeneracy between $\omega_\Lambda$ and
$\omega_m$ \cite{efs:01}. With our choice of the parameter set, $\R$ is an
independent variable, while the Hubble parameter $h$ becomes a dependent
one.

We assume throughout purely adiabatic initial conditions and
we do not allow for a tensor contribution.
In the FM approach, the likelihood distribution for the
parameters $\bf \Theta$ is expanded to quadratic order around its maximum.
We denote this maximum likelihood (ML) point by $\bf \Theta_0$ and call
the corresponding model our ``ML model'', with parameters 
$\omega_b = 0.0200$ (hence $\Omega_b = 0.0473$),
$\omega_m = 0.1267$ (hence $\Omega_m = 0.3000$),
$\omega_\Lambda = 0.2957$ (hence $\Omega_\Lambda = 0.7000$ and 
$h = 0.65$), 
$\R = 0.9628$, 
$n_s = 1.00$,
$Q = 1.00$.
For the value of $z_{\rm dec}$ in eq.~(\ref{eq:ydef})(which is
weakly dependent on $\omega_b$ and $\omega_{\rm tot}$) we
have used the fitting formula from \cite{HuandSugiyama}.
For the ML model we have $z_{\rm dec} = 1115.52$.

\begin{table}[hp]
\caption{\label{exppar} Experimental parameters for MAP and Planck 
(nominal mission). Note that the 
sensitivities are here expressed in $\mu$K.}
\begin{ruledtabular}
\begin{tabular}{|l|ccc|ccc|}
& \multicolumn{3}{c|}{MAP}& \multicolumn{3}{c|}{Planck} \\\hline
$\nu$ (GHz) &  $41$  &  $61$  & $95$ & 
               $100$ &  $143$ & $217$  \\
$\theta_c$ (arcmin)&  
	$31.8$ & $21.0$  & $13.8$ &
	$10.7$ & $8.0$ & $5.5$ \\
$\sigma_c$ ($\mu$K)  & 
	$19.8$  & $30.0$ & $45.6$ &   
	$4.6$  & $5.4$  & $11.7$  \\
$w^{-1}_c \cdot 10^{15}$ (K$^2$ ster) & 
	$33.6$  & $33.6$  & $33.6$  & 
	$0.215$ & $0.158$ & $0.350$  \\
$\ell_c$            & $254$ & $385$  & $586$  &
	$757$ & $1012$ & $1472$ \\\hline
$\ell_{\rm max}$ & \multicolumn{3}{c|}{$1000$} & \multicolumn{3}{c|}{$2000$} \\
$f_{\rm sky}$    & \multicolumn{3}{c|}{$0.80$} & \multicolumn{3}{c|}{$0.80$} \\

\end{tabular}
\end{ruledtabular}
\end{table}

Proceeding as described in \cite{Bowen:01}, 
we then calculate the {\em Fisher information matrix}
\begin{equation}
F_{ij} = \sum_{\ell=2}^{\ell_{\rm max}} \frac{1}{\Delta C_\ell^2}
	       \frac{\partial C_\ell}{\partial \Theta_i}\frac{\partial C_\ell}{\partial \Theta_j}
				 \vert_{\bf \Theta_0}
\label{eq:fisher}
\end{equation}
The quantity $\Delta C_\ell$ is the standard deviation 
on the estimate of $C_{\ell}$: 
\begin{equation}
\Delta C_\ell^2 = \frac{2}{(2 \ell + 1) f_{\rm sky} }
	               ( C_\ell + {B}_{\ell}^{-2})^2.
\end{equation}
The first term is the cosmic variance, arising from the fact
that we exchange ensemble average with a spatial average. The second
term takes into account the expected error of the experimental
apparatus \cite{knox:95,efs:01}: 

\begin{equation}
{B}_{\ell}^2 = \sum_c w_c e^{- \ell (\ell +1)/\ell_c^2}\, .
\end{equation}
The sum runs over all channels of the experiment, with
the inverse weight per solid angle
$w_c^{-1} \equiv (\sigma_c \theta_c)^{-2}$ and 
$\ell_c \equiv \sqrt{8 \ln2}/\theta_c$, where
$\sigma_c$ is the sensitivity (in $\mu$K) and 
$\theta_c$ is the FWHM of the beam (assuming a 
Gaussian profile) for each channel. Furthermore
we can neglect the issues arising from point sources,
foreground removal and galactic plane contamination
assuming that once they have been taken into account
we are left with a ``clean'' fraction of the sky given by $f_{\rm sky}$.

The experimental parameters are summarized in Table \ref{exppar}.
We use the 3 higher frequency MAP channels and the first 3 channels of the
Planck High Frequency Instrument (HFI). Adding the 3 higher frequency
channels of the HFI and the 3 channels of  Planck's Low Frequency
Instrument leaves the expected errors unchanged: therefore they can be
used for foreground removal, consistency checks, etc, leaving the HFI
channels for cosmological use.

For Gaussian fluctuations, the covariance matrix is then given by the
inverse of the Fisher matrix, $C = F^{-1}$ \cite{EBT}. 
The $1\sigma$ error on the parameter $\Theta_i$ with all other parameters
marginalised is then given by $\sqrt{C_{ii}}$. If all other parameters
are held fixed to their ML values, the standard deviation on parameter $\Theta_i$ 
reduces to $\sqrt{1/F_{ii}}$ (conditional value). 
Other cases, in which some of the parameters are
held fixed and others are being marginalized over can easily be worked out.

A case of interest is the one in which all parameters are being
estimated jointly: then the joint error on parameter $i$ is
given by the projection on the $i$-th coordinate axis of the 7-dimensional
hyper-ellipse which contains a fraction $\gamma$ of the joint likelihood.
The equation of the hyper-ellipse is
\begin{equation}
({\bf \Theta - \Theta_0}) {\bf F } ({\bf \Theta - \Theta_0})^t =
q_{1-\gamma},
\end{equation}
where $q_{1-\gamma}$ is the quantile for the probability $1-\gamma$ for
a $\chi^2$ distribution with 7 degrees of freedom. For
$\gamma = 0.683$ ($1\sigma$ c.l.) we have $q_{1-\gamma} = 8.18$.

The FMA {\it assumes} that we are expanding the 
likelihood function at the right point, ie
that the parameters values of the true model are 
in the vicinity of ${\bf \Theta_0}$. The validity
of the results depends on this assumption, as well as 
on the assumption that the $a_{\ell m}$'s are independent
Gaussian random variables. If the FM predicted errors are small
enough, the method is self-consistent and we can expect the FM prediction to
reproduce in a correct way the exact behaviour. This is indeed the case for
the present analysis, with the notable exception of $\omega_\Lambda$, which suffers
from the geometrical degeneracy (see next section).

Special care must be taken in computing the derivatives of the power spectrum
with respect to
the cosmological parameters. Numerical errors in the spectra can lead to 
larger derivatives, which would artificially break degeneracies among parameters. 
In the present work we implement double--sided derivatives, which diminish the 
truncation error from second order to third order terms. The choice of the step size
is a trade-off between truncation error and numerical inaccuracy dominated cases. 
For an estimated numerical precision of the computed models of order $10^{-4}$, 
the step size should be approximately 5\% of the parameter value \cite{Numerical:92}.
It turns out that for derivatives in direction $\alpha$ and $n_s$ 
the step size can be chosen to be as small
as $0.1\%$. As for the other parameters, the accuracy is limited by the fact that 
differentiating around a flat model requires computing open and closed models, which are
calculated using different numerical techniques.
The relative numerical noise is therefore much
larger. After several tests, we chose step sizes varying from $1\%$ to $5\%$
for $\omega_b, \omega_m, \omega_\Lambda$, and $\R$. 
This choice gives derivatives with an accuracy of about $0.5\%$. The derivatives
with respect to
$Q$ are exact, being the power spectrum itself.

\subsection{Analysis results}

\subsubsection{FMA forecast}

Table \ref{fmaresults} summarizes the results of our FMA. 
MAP will be able to constrain variations in $\alpha$ at 
the time of last scattering to
within $2\%$ ($1\sigma$, all others marginalised). This
corresponds to an improvement of a factor of 3 relative to the limits
presented in the previous section. Planck will narrow it down
to about half a percent. If all other parameters are supposed to be known
and fixed to their ML value, then a factor of 10 is to be gained in the
accuracy of $\al$ (compare the columns labelled ``fixed'' in table 
\ref{fmaresults}). However, if all parameters are being estimated
jointly, the accuracy on variations in $\alpha$ will not go beyond
$1\%$, even for Planck (column ``joint'').

The parameters $\omega_b, \R$ and $n_s$ suffer
from partial degeneracies with $\al$, which are discussed in more detail
in the next section. This is only partially reflected in the marginalized
errors of table \ref{fmaresults}. Correlations among the parameters
play an important role: within the limit of the quadratic order 
approximation, they are fully described by the FM.

The geometrical
degeneracy limits the accuracy on $\omega_m$ and $\omega_\Lambda$. The degeneracy is
so severe that the error on $\omega_\Lambda$ is very unsensitive to
the experimental details. From the FMA point of view, this happens
because the derivative of the spectrum with respect
to $\omega_\Lambda$ vanishes
for $\ell \gtrsim 50$. Therefore probing higher multipoles does not help for
the purpose of better constraining the cosmological constant. 
We emphasize once more that such a large
error cannot be trusted to be accurate in any respect: it just signals
a very large inaccuracy in $\omega_\Lambda$. The errors on all other parameters,
however, are small enough to justify the self-consistency of the FMA approach. 

\begin{table*}
\caption{\label{fmaresults}Fisher matrix analysis results: expected $1\sigma$ errors for 
the MAP and Planck satellites. The column {\it marg.} gives the error with all
other parameters being marginalized over; in the column {\it fixed} the other
parameters are held fixed at their ML value; in the column {\it joint} all
parameters are being estimated jointly.}
\begin{ruledtabular}
\begin{tabular}{|c|c c c| c c c|}
Quantity &  \multicolumn{6}{c}{$1\sigma$ errors (\%)} \\\hline  
                & \multicolumn{3}{c|}{MAP}           & \multicolumn{3}{c}{Planck HFI} \\ 
                        & marg. & fixed  & joint    & marg.      & fixed & joint\\\hline                                                       
$\al$		               &   2.24 &  0.13  & 6.39    & 0.41       & 0.02  & 1.16  \\
$\omega_b$              &   5.11 &  1.12 & 14.61    & 0.98       & 0.31 & 2.79  \\
$\omega_m$              &   5.26 &  1.97 & 15.04    & 2.30       & 0.44 & 6.59  \\ 
$\omega_\Lambda $       &   97.81&  89.62&279.74    & 95.17      & 89.55& 272.18  \\
$\R$                    &  3.73  &   0.20& 10.67    & 0.57       & 0.03 & 1.64  \\
$n_s$                   &  1.79  &   0.52&  5.12    & 1.19       & 0.13 & 3.42       \\  
$Q$                     &  1.19  &   0.36&  3.41    & 0.19       & 0.10 & 0.54 \\

\end{tabular}
\end{ruledtabular}
\end{table*}

The power of an experiment can be roughly assessed by looking at the 
eigenvalues $\lambda_i$ and eigenvectors ${\bf u^{(i)}}$ of its FM: 
the error along the direction in parameter space defined by ${\bf u^{(i)}}$ 
(principal direction) is proportional to $\lambda_i^{-1/2}$.
But we are interested in determining the errors on the physical parameters
rather then on their linear combinations along the principal directions. 
Therefore in the ideal case we want the principal directions 
to be as much aligned as possible to the coordinate system 
defined by the physical parameters. We display in Table \ref{fmaev}
eigenvalues and eigenvectors of the FM for MAP and Planck.
Planck's errors, as measured by the inverse square root of the
eigenvalues, are smaller by a factor of about 4 on average. For 6 of the 7
eigenvectors Planck also obtains a better alignment of the principal
directions with the axis of the physical parameters.
This is established by comparing the ratios between the largest (marked
with an asterisk in Table \ref{fmaev}) and the second largest (marked with
a dagger) cosmological parameters' contribution to the principal
directions.
This is of course in a slightly different
form the statement that Planck will measure the cosmological parameters
with less correlations among them.

\begin{table*}
\caption{\label{fmaev} In the lines we display the components
of the eigenvectors of the FM for MAP and
Planck. The quantity $1/\sqrt{\lambda_i}$ is proportional to the error along
the principal direction ${\bf u^{(i)}}$. For each principal direction, an
asterisk marks the largest cosmological parameters contribution, a dagger
the second largest. Not only Planck has errors smaller by a factor of about
4 on average, but also the alignment of the principal directions with the
axis defined by the physical parameter is better than MAP in 6 cases out of 7.}
\begin{ruledtabular}
\begin{tabular}{|c c c c c c c c c|}
\multicolumn{9}{|c|}{MAP} \\\hline
Direction $i$ & $1/\sqrt{\lambda_i}$ & $\omega_b$ & $\omega_m$ &
$\omega_\Lambda$ & $n_s$ & $Q$ & $\R$ & $\alpha$ \\\hline
1& 2.23E-04 & 9.9540E-01* & -9.7492E-03 & -1.8938E-05 & -1.6970E-02
& -4.2258E-03 & -6.8163E-02$\dag$ & 6.4257E-02 \\
2& 1.12E-03 & -7.8215E-02 & 2.8225E-01 & 1.6168E-05 & 1.6425E-01
& -1.4652E-01 & -4.4305E-01$\dag$& 8.1822E-01* \\
3& 2.43E-03 & 4.0471E-02 & 7.2825E-01* & 2.0996E-04 & 1.4674E-01
& -5.6711E-01$\dag$& 2.6310E-01 & -2.3590E-01 \\
4&8.63E-03 & 3.9968E-03 & 6.1138E-01$\dag$& 8.1457E-03 & -3.7196E-01 &
6.4228E-01* & -2.3162E-01 & -1.4702E-01 \\
5& 4.45E-02 &-1.9844E-02 & -6.8856E-02 & -1.6683E-02 & 2.5892E-01
& -1.8786E-01 & -8.0247E-01* & -4.9829E-01$\dag$\\
6& 1.38E-02& 3.1792E-02 & 1.0650E-01 & -1.5423E-02 & 8.6336E-01* &
4.5726E-01$\dag$& 1.7932E-01 & -2.8026E-02 \\
7& 2.89E-01 & 2.0351E-04 & -4.6453E-03 & 9.9971E-01* &
2.0638E-02$\dag$& -1.1925E-03 & -8.7876E-03 & -7.5125E-03 \\\hline
\multicolumn{9}{|c|}{Planck} \\\hline
Direction $i$ & $1/\sqrt{\lambda_i}$ & $\omega_b$ & $\omega_m$ &
$\omega_\Lambda$ & $n_s$ & $Q$ & $\R$ & $\alpha$ \\\hline
1& 5.81E-05 & 9.3678E-01* & 8.2153E-03 & -1.2417E-06 & 9.2222E-03 &
3.0060E-03 & -1.8611E-01 & 2.9604E-01$\dag$\\
2& 2.87E-04 & -3.2830E-01 & 2.5649E-01 & 9.3971E-06 & 1.5805E-01 &
2.0156E-01 & -3.8542E-01$\dag$& 7.8248E-01* \\
3& 5.46E-04 & 1.1323E-01 & 8.4059E-01* & 4.5325E-06 & 2.7521E-01 &
2.5135E-01 & 3.2628E-01$\dag$& -1.8765E-01 \\
4&1.94E-03 & 3.8999E-02 & -3.1741E-01$\dag$ & -9.3635E-05 & -3.8346E-02 &
9.4091E-01* & 6.4234E-02 & -8.2576E-02 \\
5& 2.86E-03 & -1.5404E-02 & 2.9918E-01 & 7.3411E-04 & -3.9324E-01 &
9.9942E-02 & -7.5339E-01* & -4.2194E-01$\dag$ \\
6& 1.30E-02 & 8.7278E-03 & -1.9309E-01 & -1.6271E-02 & 8.6190E-01*
& -2.9783E-02 & -3.7230E-01$\dag$& -2.8285E-01 \\
7& 2.81E-01 & 1.6073E-04 & -3.3978E-03 & 9.9987E-01* &
1.4308E-02$\dag$& -4.7295E-04 & -5.4974E-03 & -4.3069E-03 \\
\end{tabular}
\end{ruledtabular}
\end{table*}

\subsubsection{Degeneracies with other parameters}

In previous work \cite{Avelino:2000ea} some of the present authors observed a 
degeneracy in the Boomerang-98 and Maxima-1 data between $\al$ and $\omega_b$. 
This degeneracy also shows up in the
present analysis (see Fig.  \ref{fig1}, top panel). The question we ask is: 
is this a fundamental degeneracy, or is it only in the data? 

\begin{figure}
\includegraphics[width=3in,keepaspectratio]{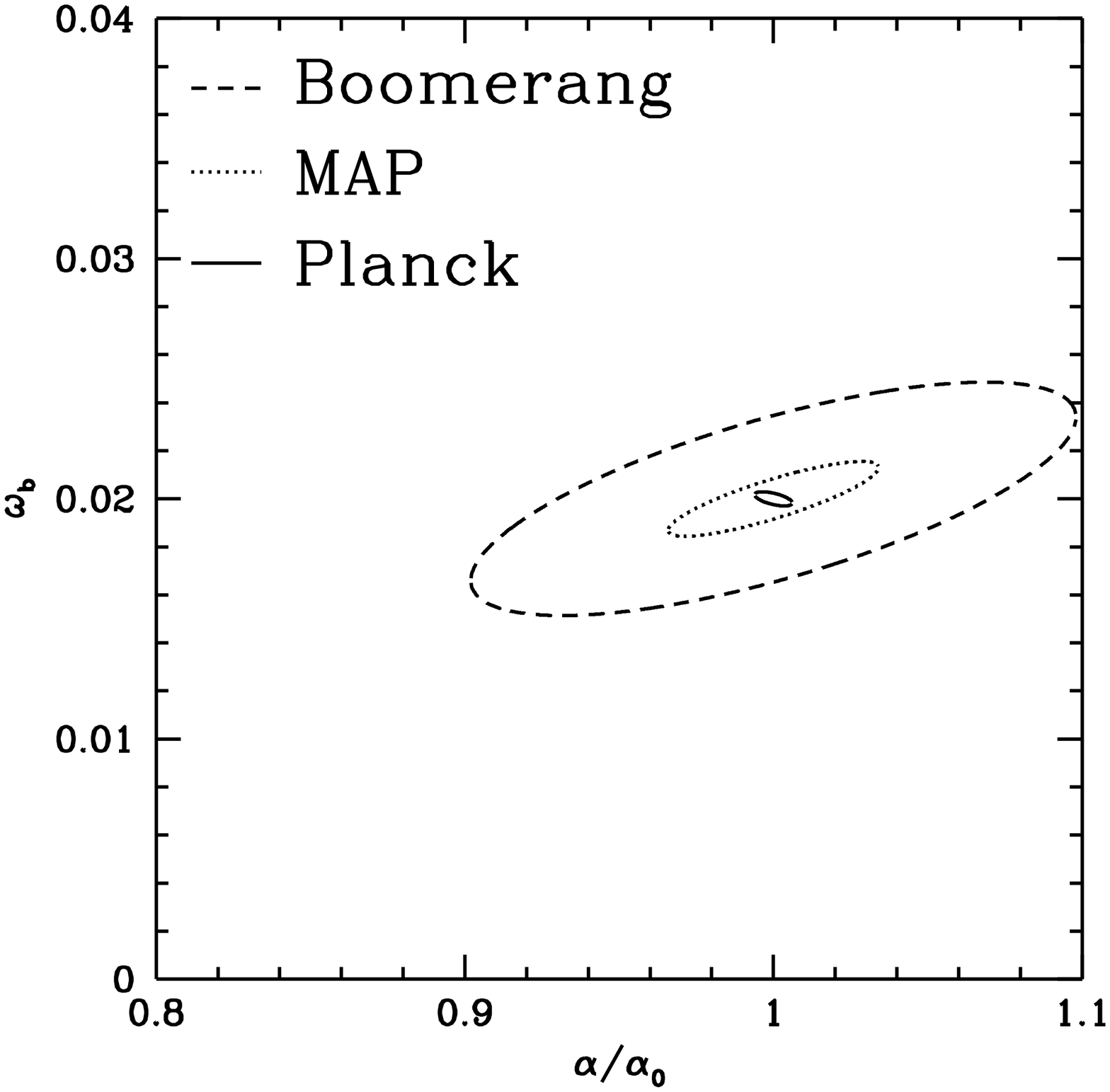}
\includegraphics[width=3in,keepaspectratio]{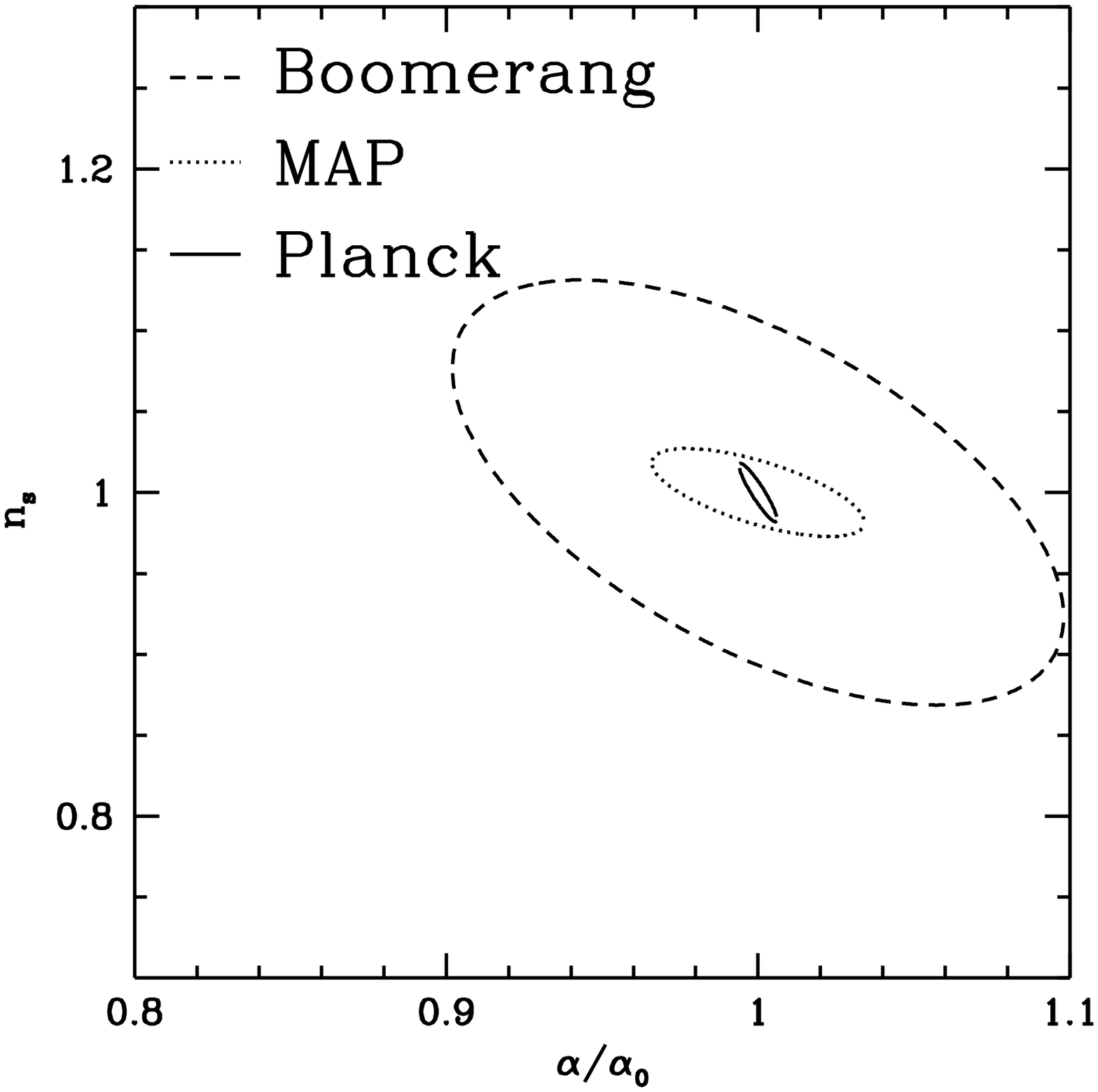}
\includegraphics[width=3in,keepaspectratio]{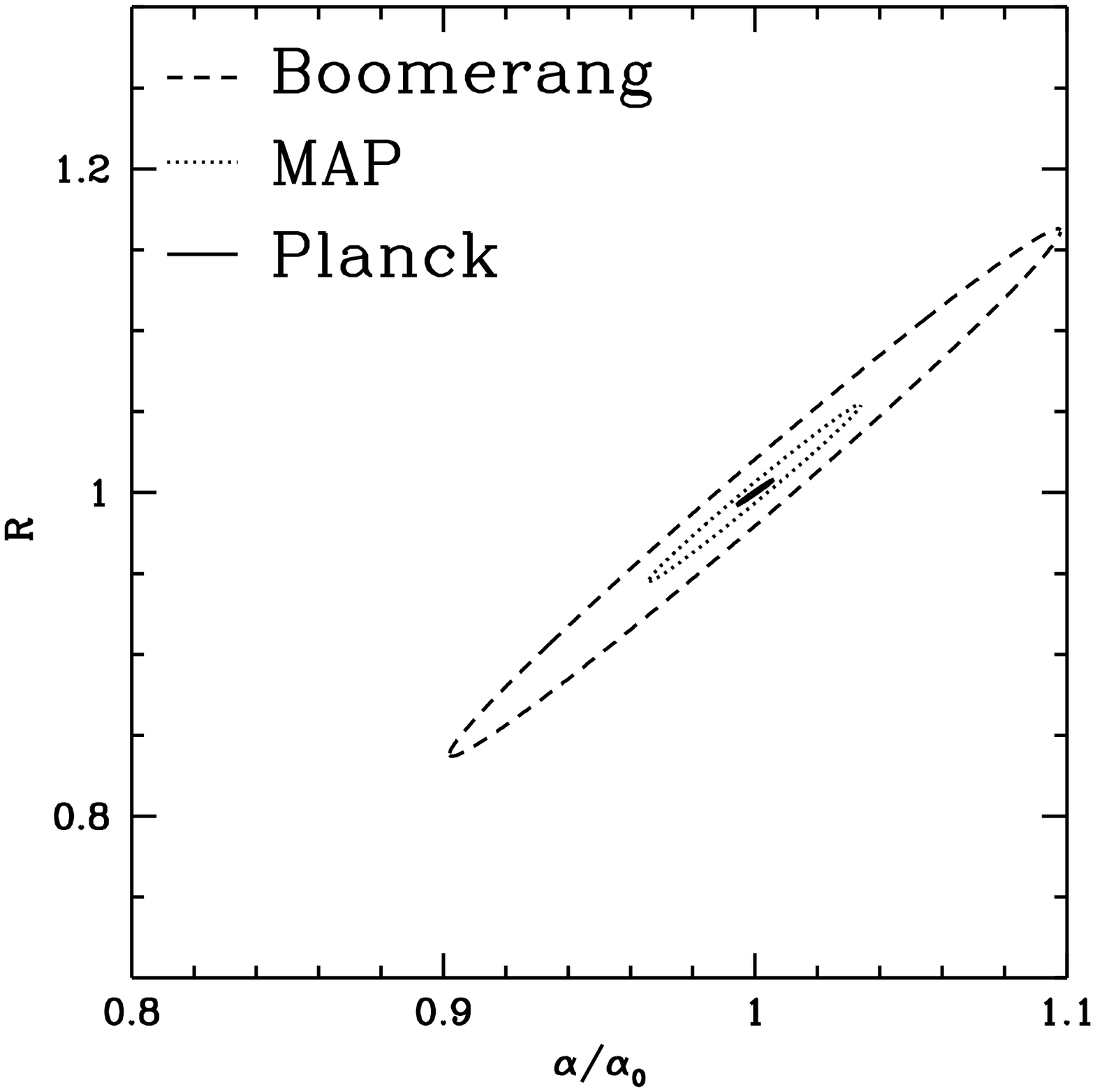}
\caption{\label{figellipse} Ellipses containing $68\%$ ($1\sigma$) of
joint confidence (all other
parameters marginalized) in the $\al /\alpha_0 - \omega_b$,
$\al /\alpha_0- n_s$ and
$\al /\alpha_0- \R$ planes for the FM simulated
Boomerang, MAP and Planck.}  
\end{figure}

We have performed a FMA with experimental parameters chosen as to
mimic a Boomerang-type experiment ($f_{\rm sky} = 0.02, 
\sigma = 15 \quad \mu\rm{K},
\theta = 9.2', 
\ell_{\rm max} = 1000$). If the degeneracy is due to
the limited precision of the present-day experimental data, we expect the
degeneracy to disappear as we move from Boomerang, to MAP, to Planck. 
Fig. \ref{figellipse} (top panel) shows $1\sigma$ joint confidence
curves (all other
parameters marginalized) in the $\al /\alpha_0- \omega_b$
plane for the FM simulated
Boomerang, MAP and Planck (from the outside to the center, respectively). 
The curve for Boomerang is to be compared with the $1\sigma$ contour of the
data analysis (Figure \ref{fig1}, top panel). Although the FM ellipse is centred
by construction at the ML model value, it is in qualitative agreement
with the result of the data analysis. As we move to MAP, the degeneracy shrinks
but is still there: only higher multipole measurements from Planck can break
it.

The same behaviour is observed in the $\al /\alpha_0- n_s$ plane
(Fig. \ref{figellipse},
middle panel;  compare with Fig. \ref{fig1}, middle panel).
Again, the observed degeneracy between
$\al$ and $n_s$ is clearly revealed by the FMA for Boomerang.

\begin{figure}
\includegraphics[width=3in,keepaspectratio]{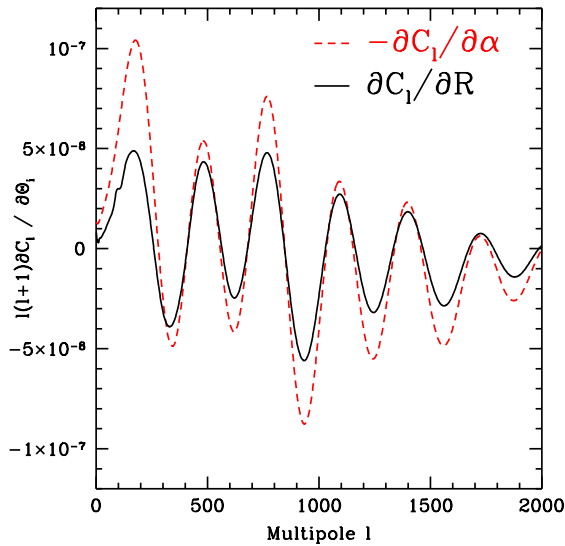}
\caption{\label{figderalR} Derivatives with respect to $\al$ and
$\R$. We plot $-\partial C_\ell/ \partial \al$ to facilitate
the comparison with $\partial C_\ell/ \partial \R$. The two
derivatives are perfectly in phase: this is responsible
for the degeneracy between the corresponding parameters. Only
the different amplitudes allow an experiment which maps 
sufficiently high multipoles with high accuracy to distinguish between them.}
\end{figure}

In the bottom panel
of Fig. \ref{figellipse} we investigate the important
degeneracy between $\R$ and $\al /\alpha_0$. These
two parameters are very highly correlated (correlation $\approx 0.99$ for all
experiments), because an increase of $\alpha$ displaces the acoustic peaks 
to higher multipoles. This effect is mainly due to the increased redshift 
of last scattering \cite{Kaplinghat:99,Avelino:2000ea}. On the contrary, 
an increase of $\R$ shifts the peaks toward smaller $\ell$ values, 
because of the change in the angular diameter distance relation. However,
an increase in $\al$ also produces a decrease in the damping at high
multipoles, which can
be used to break the degeneracy, as it is the case for Planck. This is
reflected in the different amplitude for the two derivatives, which
are otherwise perfectly in phase, as can be seen in Fig. \ref{figderalR}.
This degeneracy can 
clearly be identified because of our choice of the parameter set
${\bf \Theta}$, which
includes the shift parameter rather than the Hubble or curvature
parameters. This emphasizes the importance of a correct choice of 
the parameter set in the context of a FMA.

Comparison to previous works is only partially possible, because
of the differences in the analysis discussed above. Our detailed
analysis confirms however the conclusions in refs. 
\cite{Hannestad:99,Kaplinghat:99}, which found that a cosmic 
variance limited experiment could obtain a precision on $\alpha$
of order $10^{-3} - 10^{-2}$. We have also shown that there is much
to be gained from using prior knowledge about the other
parameters in the determination of $\alpha$ via CMB
measurements. The improvement in accuracy is about a factor
$50$ for both MAP and Planck. 
 
\section{\label{conc}Conclusions}

We have provided an up-to-date analysis of the effects of
a varying fine-structure constant $\alpha$ in the CMB, focusing
on the issue of the degeneracies with other cosmological parameters,
and of how these can be broken.

We have shown that the currently available data is consistent with no
variation of $\alpha$ from the epoch of recombination to the
present day, though interestingly enough the CMB and LSS
datasets seem to prefer, on their own, variations of $\alpha$ with
opposite signs. Whether or not this statement has any physical
relevance (beyond the results of the statistical analysis) is
something that remains to be investigated in more detail. In any case,
any such (relative) variation is constrained to be less than
about $4\%$, so a best-fit or `concordance' model with $\alpha$ exactly
constant will require, at most, some slight deviations of other
cosmological parameters from the `standard' values obtained from
analyses which don't allow for $\alpha$ variation.

On the other hand, the prospects for the future are definitely bright.
In the short term, the imminent VSA and CBI data should be able to
provide some improvement
on the current results. The dataset we have used (BOOMERanG, MAXIMA
and DASI) all have the common feature that their error bars are smallest
for data points around the first Doppler peak and larger for smallest
angular scales. Now, as we have explicitly shown above (and was already
suggested in \cite{Avelino:2000ea}), the first Doppler peak is not a
vary accurate `$\alpha$-meter', due to the degeneracy with the shift
parameter. Thus datasets in which points around the first peak
will somewhat dominate the statistical analysis are not optimal for
$\alpha$ estimation. In this regard, VSA and CBI should be useful because
they can provide a significant number of data points on small angular
scales with relatively small error bars, hence minimizing this problem.

In the longer term, the forthcoming satellite experiments will provide
a dramatic improvement on these results.
We have performed a Fisher Matrix Analysis using a well adapted
parameter set and realistic experimental characteristics for the upcoming
MAP and Planck satellite missions. The results of our forecast are 
that MAP and Planck will be able
to constrain variations in $\alpha$ within $2.2\%$ and $0.4\%$, respectively
($1\sigma$ c.l., all other parameters marginalized). If all parameters are
being estimated simultaneously, then this limits increase to about
$6.4\%$ and $1.1\%$, respectively. The analysis of the presently observed
degeneracies between $\alpha$ and $\omega_b$, $n_s$ comes to the conclusion
that measurement of higher multipoles will allow to break it. We have
also identified an important degeneracy between $\alpha$ and the shift
parameter.

To conclude, we have provided a thorough analysis of the effects of
cosmological parameter degeneracies in CMB measurements of the
fine-structure constant $\alpha$, and quantified the importance of
these degeneracies. We have also explicitly discussed two ways in
which these degeneracies can be circumvented, namely acquiring
better data (the easy solution, at least from the theorists' point of view)
or combining the CMB data with other cosmological datasets which can
provide constraints on other cosmological parameters (the `brute-force' solution)

In a follow-up paper, we will discuss a third way in which 
these degeneracies can be lifted, namely including CMB polarization
data \cite{Rocha:2002} (the more elegant solution in principle, though it's
yet to be realized in practice). These tools, together with other
measurements coming from BBN \cite{Avelino:2000ea} and quasar and
related data \cite{Murphy:2000pz,Webb:2000mn} offer the exciting prospect
of being able to map the value of $\alpha$ at very many different cosmological
epochs, which would allow us to impose very tight constraints on
higher-dimensional models where these variations are ubiquitous.

\begin{acknowledgments}
    
We are grateful to Ruth Durrer, Stefano Foffa, Yasmin Friedmann,
Arthur Kosowsky, Lyman A. Page, 
Paul Shellard, David N. Spergel, Carsten van de Bruck and John Webb
for many useful discussions.

This work is partially supported by the European Network CMBNET.
C.M. is funded by FCT (Portugal), under grant no. FMRH/BPD/1600/2000.
A.M. and R.B. are supported by PPARC.
R.T. is partially supported by the Schmidheiny Fundation.
G.R. is funded by a Leverhulme Fellowship.

This work was performed on COSMOS, the Origin2000 owned by the UK
Computational Cosmology Consortium, supported by Silicon Graphics/Cray
Research, HEFCE and PPARC.

\end{acknowledgments}

\bibliography{alpha_low}

\begin{thebibliography}{45}
\expandafter\ifx\csname natexlab\endcsname\relax\def\natexlab#1{#1}\fi
\expandafter\ifx\csname bibnamefont\endcsname\relax
  \def\bibnamefont#1{#1}\fi
\expandafter\ifx\csname bibfnamefont\endcsname\relax
  \def\bibfnamefont#1{#1}\fi
\expandafter\ifx\csname citenamefont\endcsname\relax
  \def\citenamefont#1{#1}\fi
\expandafter\ifx\csname url\endcsname\relax
  \def\url#1{\texttt{#1}}\fi
\expandafter\ifx\csname urlprefix\endcsname\relax\def\urlprefix{URL }\fi
\providecommand{\bibinfo}[2]{#2}
\providecommand{\eprint}[2][]{\url{#2}}

\bibitem[{\citenamefont{Murphy et~al.}(2001{\natexlab{a}})}]{Murphy:2000pz}
\bibinfo{author}{\bibfnamefont{M.~T.} \bibnamefont{Murphy}}
  \bibnamefont{et~al.}, \bibinfo{journal}{Mon. Not. Roy. Astron. Soc.}
  \textbf{\bibinfo{volume}{327}}, \bibinfo{pages}{1208}
  (\bibinfo{year}{2001}{\natexlab{a}}), \eprint{arXiv:astro-ph/0012419}.

\bibitem[{\citenamefont{Webb et~al.}(2001)}]{Webb:2000mn}
\bibinfo{author}{\bibfnamefont{J.~K.} \bibnamefont{Webb}} \bibnamefont{et~al.},
  \bibinfo{journal}{Phys. Rev. Lett.} \textbf{\bibinfo{volume}{87}},
  \bibinfo{pages}{091301} (\bibinfo{year}{2001}),
  \eprint{arXiv:astro-ph/0012539}.

\bibitem[{\citenamefont{Murphy et~al.}(2001{\natexlab{b}})\citenamefont{Murphy,
  Webb, Flambaum, Prochaska, and Wolfe}}]{Murphy:2000ns}
\bibinfo{author}{\bibfnamefont{M.~T.} \bibnamefont{Murphy}},
  \bibinfo{author}{\bibfnamefont{J.~K.} \bibnamefont{Webb}},
  \bibinfo{author}{\bibfnamefont{V.~V.} \bibnamefont{Flambaum}},
  \bibinfo{author}{\bibfnamefont{J.~X.} \bibnamefont{Prochaska}},
  \bibnamefont{and} \bibinfo{author}{\bibfnamefont{A.~M.} \bibnamefont{Wolfe}},
  \bibinfo{journal}{Mon. Not. Roy. Astron. Soc.}
  \textbf{\bibinfo{volume}{327}}, \bibinfo{pages}{1237}
  (\bibinfo{year}{2001}{\natexlab{b}}), \eprint{arXiv:astro-ph/0012421}.

\bibitem[{\citenamefont{Murphy et~al.}(2001{\natexlab{c}})}]{Murphy:2001nu}
\bibinfo{author}{\bibfnamefont{M.~T.} \bibnamefont{Murphy}}
  \bibnamefont{et~al.}, \bibinfo{journal}{Mon. Not. Roy. Astron. Soc.}
  \textbf{\bibinfo{volume}{327}}, \bibinfo{pages}{1244}
  (\bibinfo{year}{2001}{\natexlab{c}}), \eprint{arXiv:astro-ph/0101519}.

\bibitem[{\citenamefont{Avelino et~al.}(2001)}]{Avelino:2001nr}
\bibinfo{author}{\bibfnamefont{P.~P.} \bibnamefont{Avelino}}
  \bibnamefont{et~al.}, \bibinfo{journal}{Phys. Rev.}
  \textbf{\bibinfo{volume}{D64}}, \bibinfo{pages}{103505}
  (\bibinfo{year}{2001}), \eprint{arXiv:astro-ph/0102144}.

\bibitem[{\citenamefont{Fujii}(2002)}]{Fujii:2002bi}
\bibinfo{author}{\bibfnamefont{Y.}~\bibnamefont{Fujii}} (\bibinfo{year}{2002}),
  \eprint[http://arXiv.org/abs]{astro-ph/0204069}.

\bibitem[{\citenamefont{Ivanchik et~al.}(2001)}]{Ivanchik:2001ji}
\bibinfo{author}{\bibfnamefont{A.}~\bibnamefont{Ivanchik}} \bibnamefont{et~al.}
  (\bibinfo{year}{2001}), \eprint{arXiv:astro-ph/0112323}.

\bibitem[{\citenamefont{Avelino
  et~al.}(2000{\natexlab{a}})\citenamefont{Avelino, Martins, and
  Rocha}}]{Avelino:2000oo}
\bibinfo{author}{\bibfnamefont{P.}~\bibnamefont{Avelino}},
  \bibinfo{author}{\bibfnamefont{C.}~\bibnamefont{Martins}}, \bibnamefont{and}
  \bibinfo{author}{\bibfnamefont{G.}~\bibnamefont{Rocha}},
  \bibinfo{journal}{Phys. Lett.} \textbf{\bibinfo{volume}{B483}},
  \bibinfo{pages}{210} (\bibinfo{year}{2000}{\natexlab{a}}),
  \eprint{arXiv:astro-ph/0001292}.

\bibitem[{\citenamefont{Avelino
  et~al.}(2000{\natexlab{b}})\citenamefont{Avelino, Martins, Rocha, and
  Viana}}]{Avelino:2000ea}
\bibinfo{author}{\bibfnamefont{P.~P.} \bibnamefont{Avelino}},
  \bibinfo{author}{\bibfnamefont{C.~J. A.~P.} \bibnamefont{Martins}},
  \bibinfo{author}{\bibfnamefont{G.}~\bibnamefont{Rocha}}, \bibnamefont{and}
  \bibinfo{author}{\bibfnamefont{P.}~\bibnamefont{Viana}},
  \bibinfo{journal}{Phys. Rev.} \textbf{\bibinfo{volume}{D62}},
  \bibinfo{pages}{123508} (\bibinfo{year}{2000}{\natexlab{b}}),
  \eprint{arXiv:astro-ph/0008446}.

\bibitem[{\citenamefont{Battye et~al.}(2001)\citenamefont{Battye, Crittenden,
  and Weller}}]{Battye:2000ds}
\bibinfo{author}{\bibfnamefont{R.~A.} \bibnamefont{Battye}},
  \bibinfo{author}{\bibfnamefont{R.}~\bibnamefont{Crittenden}},
  \bibnamefont{and} \bibinfo{author}{\bibfnamefont{J.}~\bibnamefont{Weller}},
  \bibinfo{journal}{Phys. Rev.} \textbf{\bibinfo{volume}{D63}},
  \bibinfo{pages}{043505} (\bibinfo{year}{2001}),
  \eprint{arXiv:astro-ph/0008265}.

\bibitem[{\citenamefont{Huey et~al.}(2002)\citenamefont{Huey, Alexander, and
  Pogosian}}]{Huey:2001ku}
\bibinfo{author}{\bibfnamefont{G.}~\bibnamefont{Huey}},
  \bibinfo{author}{\bibfnamefont{S.}~\bibnamefont{Alexander}},
  \bibnamefont{and} \bibinfo{author}{\bibfnamefont{L.}~\bibnamefont{Pogosian}},
  \bibinfo{journal}{Phys. Rev.} \textbf{\bibinfo{volume}{D65}},
  \bibinfo{pages}{083001} (\bibinfo{year}{2002}),
  \eprint{arXiv:astro-ph/0110562}.

\bibitem[{\citenamefont{Rocha et~al.}(2002)}]{Rocha:2002}
\bibinfo{author}{\bibfnamefont{G.}~\bibnamefont{Rocha}} \bibnamefont{et~al.}
  (\bibinfo{year}{2002}), \bibinfo{note}{in preparation}.

\bibitem[{\citenamefont{Webb}(2001)}]{Webb:2001}
\bibinfo{author}{\bibfnamefont{J.~K.} \bibnamefont{Webb}}
  (\bibinfo{year}{2001}), \bibinfo{note}{private communication}.

\bibitem[{\citenamefont{Damour and Dyson}(1996)}]{Damour:1996zw}
\bibinfo{author}{\bibfnamefont{T.}~\bibnamefont{Damour}} \bibnamefont{and}
  \bibinfo{author}{\bibfnamefont{F.}~\bibnamefont{Dyson}},
  \bibinfo{journal}{Nucl. Phys.} \textbf{\bibinfo{volume}{B480}},
  \bibinfo{pages}{37} (\bibinfo{year}{1996}), \eprint{arXiv:hep-ph/9606486}.

\bibitem[{\citenamefont{Fujii et~al.}(2000)}]{Fujii:1998kn}
\bibinfo{author}{\bibfnamefont{Y.}~\bibnamefont{Fujii}} \bibnamefont{et~al.},
  \bibinfo{journal}{Nucl. Phys.} \textbf{\bibinfo{volume}{B573}},
  \bibinfo{pages}{377} (\bibinfo{year}{2000}), \eprint{arXiv:hep-ph/9809549}.

\bibitem[{\citenamefont{Bekenstein}(1982)}]{Bekenstein:1982eu}
\bibinfo{author}{\bibfnamefont{J.~D.} \bibnamefont{Bekenstein}},
  \bibinfo{journal}{Phys. Rev.} \textbf{\bibinfo{volume}{D25}},
  \bibinfo{pages}{1527} (\bibinfo{year}{1982}).

\bibitem[{\citenamefont{Sandvik et~al.}(2002)\citenamefont{Sandvik, Barrow, and
  Magueijo}}]{Sandvik:2001rv}
\bibinfo{author}{\bibfnamefont{H.~B.} \bibnamefont{Sandvik}},
  \bibinfo{author}{\bibfnamefont{J.~D.} \bibnamefont{Barrow}},
  \bibnamefont{and} \bibinfo{author}{\bibfnamefont{J.}~\bibnamefont{Magueijo}},
  \bibinfo{journal}{Phys. Rev. Lett.} \textbf{\bibinfo{volume}{88}},
  \bibinfo{pages}{031302} (\bibinfo{year}{2002}),
  \eprint[http://arXiv.org/abs]{astro-ph/0107512}.

\bibitem[{\citenamefont{Olive and Pospelov}(2002)}]{Olive:2001vz}
\bibinfo{author}{\bibfnamefont{K.~A.} \bibnamefont{Olive}} \bibnamefont{and}
  \bibinfo{author}{\bibfnamefont{M.}~\bibnamefont{Pospelov}},
  \bibinfo{journal}{Phys. Rev.} \textbf{\bibinfo{volume}{D65}},
  \bibinfo{pages}{085044} (\bibinfo{year}{2002}),
  \eprint[http://arXiv.org/abs]{hep-ph/0110377}.

\bibitem[{\citenamefont{Damour and Nordtvedt}(1993)}]{Damour:1993id}
\bibinfo{author}{\bibfnamefont{T.}~\bibnamefont{Damour}} \bibnamefont{and}
  \bibinfo{author}{\bibfnamefont{K.}~\bibnamefont{Nordtvedt}},
  \bibinfo{journal}{Phys. Rev.} \textbf{\bibinfo{volume}{D48}},
  \bibinfo{pages}{3436} (\bibinfo{year}{1993}).

\bibitem[{\citenamefont{Santiago et~al.}(1998)\citenamefont{Santiago, Kalligas,
  and Wagoner}}]{Santiago:1998ae}
\bibinfo{author}{\bibfnamefont{D.~I.} \bibnamefont{Santiago}},
  \bibinfo{author}{\bibfnamefont{D.}~\bibnamefont{Kalligas}}, \bibnamefont{and}
  \bibinfo{author}{\bibfnamefont{R.~V.} \bibnamefont{Wagoner}},
  \bibinfo{journal}{Phys. Rev.} \textbf{\bibinfo{volume}{D58}},
  \bibinfo{pages}{124005} (\bibinfo{year}{1998}),
  \eprint[http://arXiv.org/abs]{gr-qc/9805044}.

\bibitem[{\citenamefont{Braxmaier et~al.}(2001)}]{Braxmaier:2001ph}
\bibinfo{author}{\bibfnamefont{C.}~\bibnamefont{Braxmaier}}
  \bibnamefont{et~al.}, \bibinfo{journal}{Phys. Rev.}
  \textbf{\bibinfo{volume}{D64}}, \bibinfo{pages}{042001}
  (\bibinfo{year}{2001}).

\bibitem[{\citenamefont{Coc et~al.}(2002)\citenamefont{Coc, Vangioni-Flam,
  Casse, and Rabiet}}]{Coc:2002tr}
\bibinfo{author}{\bibfnamefont{A.}~\bibnamefont{Coc}},
  \bibinfo{author}{\bibfnamefont{E.}~\bibnamefont{Vangioni-Flam}},
  \bibinfo{author}{\bibfnamefont{M.}~\bibnamefont{Casse}}, \bibnamefont{and}
  \bibinfo{author}{\bibfnamefont{M.}~\bibnamefont{Rabiet}},
  \bibinfo{journal}{Phys. Rev.} \textbf{\bibinfo{volume}{D65}},
  \bibinfo{pages}{043510} (\bibinfo{year}{2002}),
  \eprint{arXiv:astro-ph/0111077}.

\bibitem[{\citenamefont{Cyburt et~al.}(2001)\citenamefont{Cyburt, Fields, and
  Olive}}]{Cyburt:2001pp}
\bibinfo{author}{\bibfnamefont{R.~H.} \bibnamefont{Cyburt}},
  \bibinfo{author}{\bibfnamefont{B.~D.} \bibnamefont{Fields}},
  \bibnamefont{and} \bibinfo{author}{\bibfnamefont{K.~A.} \bibnamefont{Olive}},
  \bibinfo{journal}{New Astron.} \textbf{\bibinfo{volume}{6}},
  \bibinfo{pages}{215} (\bibinfo{year}{2001}), \eprint{arXiv:astro-ph/0102179}.

\bibitem[{\citenamefont{Avelino and Martins}(1999)}]{Avelino:1999is}
\bibinfo{author}{\bibfnamefont{P.~P.} \bibnamefont{Avelino}} \bibnamefont{and}
  \bibinfo{author}{\bibfnamefont{C.~J. A.~P.} \bibnamefont{Martins}},
  \bibinfo{journal}{Phys. Lett.} \textbf{\bibinfo{volume}{B459}},
  \bibinfo{pages}{468} (\bibinfo{year}{1999}), \eprint{arXiv:astro-ph/9906117}.

\bibitem[{\citenamefont{Seljak and Zaldarriaga}(1996)}]{sz}
\bibinfo{author}{\bibfnamefont{U.}~\bibnamefont{Seljak}} \bibnamefont{and}
  \bibinfo{author}{\bibfnamefont{M.}~\bibnamefont{Zaldarriaga}},
  \bibinfo{journal}{Astrophys. J.} \textbf{\bibinfo{volume}{469}},
  \bibinfo{pages}{437} (\bibinfo{year}{1996}), \eprint{arXiv:astro-ph/9603033}.

\bibitem[{\citenamefont{Gnedin}(2001)}]{gnedin}
\bibinfo{author}{\bibfnamefont{N.}~\bibnamefont{Gnedin}}
  (\bibinfo{year}{2001}), \eprint{arXiv:astro-ph/0110290}.

\bibitem[{\citenamefont{Bridle et~al.}(2001)}]{sara}
\bibinfo{author}{\bibfnamefont{S.~L.} \bibnamefont{Bridle}}
  \bibnamefont{et~al.} (\bibinfo{year}{2001}), \eprint{arXiv:astro-ph/0112114}.

\bibitem[{\citenamefont{Tegmark et~al.}(2001)\citenamefont{Tegmark, Hamilton,
  and Shu}}]{thx}
\bibinfo{author}{\bibfnamefont{M.}~\bibnamefont{Tegmark}},
  \bibinfo{author}{\bibfnamefont{A.}~\bibnamefont{Hamilton}}, \bibnamefont{and}
  \bibinfo{author}{\bibfnamefont{Y.}~\bibnamefont{Shu}} (\bibinfo{year}{2001}),
  \eprint{arXiv:astro-ph/0111575}.

\bibitem[{\citenamefont{Pierpaoli et~al.}(2001)\citenamefont{Pierpaoli, Scott,
  and White}}]{pierpa}
\bibinfo{author}{\bibfnamefont{E.}~\bibnamefont{Pierpaoli}},
  \bibinfo{author}{\bibfnamefont{D.}~\bibnamefont{Scott}}, \bibnamefont{and}
  \bibinfo{author}{\bibfnamefont{M.}~\bibnamefont{White}},
  \bibinfo{journal}{Mon. Not. Roy. Astron. Soc.}
  \textbf{\bibinfo{volume}{325}}, \bibinfo{pages}{77} (\bibinfo{year}{2001}).

\bibitem[{\citenamefont{Eke et~al.}(1996)\citenamefont{Eke, Cole, and
  Frenk}}]{eke}
\bibinfo{author}{\bibfnamefont{V.~R.} \bibnamefont{Eke}},
  \bibinfo{author}{\bibfnamefont{S.}~\bibnamefont{Cole}}, \bibnamefont{and}
  \bibinfo{author}{\bibfnamefont{C.}~\bibnamefont{Frenk}},
  \bibinfo{journal}{Mon. Not. Roy. Astron. Soc.}
  \textbf{\bibinfo{volume}{282}}, \bibinfo{pages}{263} (\bibinfo{year}{1996}).

\bibitem[{\citenamefont{Seljak}(2001)}]{s8eljak}
\bibinfo{author}{\bibfnamefont{U.}~\bibnamefont{Seljak}}
  (\bibinfo{year}{2001}), \eprint{arXiv:astro-ph/0111362}.

\bibitem[{\citenamefont{Viana et~al.}(2001)\citenamefont{Viana, Nichol, and
  Liddle}}]{liddle}
\bibinfo{author}{\bibfnamefont{P.}~\bibnamefont{Viana}},
  \bibinfo{author}{\bibfnamefont{R.}~\bibnamefont{Nichol}}, \bibnamefont{and}
  \bibinfo{author}{\bibfnamefont{A.}~\bibnamefont{Liddle}}
  (\bibinfo{year}{2001}), \eprint{arXiv:astro-ph/0111394}.

\bibitem[{\citenamefont{Melchiorri et~al.}(2000)}]{melchiorri}
\bibinfo{author}{\bibfnamefont{A.}~\bibnamefont{Melchiorri}}
  \bibnamefont{et~al.}, \bibinfo{journal}{Astrophys. J.}
  \textbf{\bibinfo{volume}{536}}, \bibinfo{pages}{L63} (\bibinfo{year}{2000}),
  \eprint{arXiv:astro-ph/9911445}.

\bibitem[{\citenamefont{Stompor}(2001)}]{stompor}
\bibinfo{author}{\bibfnamefont{R.}~\bibnamefont{Stompor}},
  \bibinfo{journal}{Astrophys. J.} \textbf{\bibinfo{volume}{561}},
  \bibinfo{pages}{L7} (\bibinfo{year}{2001}), \eprint{arXiv:astro-ph/0105062}.

\bibitem[{\citenamefont{MAP}(2002)}]{MAP}
\bibinfo{author}{\bibnamefont{MAP}} (\bibinfo{year}{2002}),
  \bibinfo{note}{http://map.gsfc.nasa.gov/}.

\bibitem[{\citenamefont{Planck}(2002)}]{Planck}
\bibinfo{author}{\bibnamefont{Planck}} (\bibinfo{year}{2002}),
  \bibinfo{note}{http://astro.estec.esa.nl/Planck}.

\bibitem[{\citenamefont{Hannestad}(1999)}]{Hannestad:99}
\bibinfo{author}{\bibfnamefont{S.}~\bibnamefont{Hannestad}},
  \bibinfo{journal}{Phys. Rev. D} \textbf{\bibinfo{volume}{60}},
  \bibinfo{pages}{023515} (\bibinfo{year}{1999}),
  \eprint{arXiv:astro-ph/9810102}.

\bibitem[{\citenamefont{Kaplinghat et~al.}(1999)\citenamefont{Kaplinghat,
  Scherrer, and Turner}}]{Kaplinghat:99}
\bibinfo{author}{\bibfnamefont{M.}~\bibnamefont{Kaplinghat}},
  \bibinfo{author}{\bibfnamefont{R.}~\bibnamefont{Scherrer}}, \bibnamefont{and}
  \bibinfo{author}{\bibfnamefont{M.}~\bibnamefont{Turner}},
  \bibinfo{journal}{Phys. Rev. D} \textbf{\bibinfo{volume}{60}},
  \bibinfo{pages}{023516} (\bibinfo{year}{1999}),
  \eprint{arXiv:astro-ph/9810133}.

\bibitem[{\citenamefont{Melchiorri and Griffiths}(2001)}]{melch:01}
\bibinfo{author}{\bibfnamefont{A.}~\bibnamefont{Melchiorri}} \bibnamefont{and}
  \bibinfo{author}{\bibfnamefont{L.~M.} \bibnamefont{Griffiths}},
  \bibinfo{journal}{New Astronomy Reviews} \textbf{\bibinfo{volume}{45}},
  \bibinfo{pages}{Issue 4} (\bibinfo{year}{2001}),
  \eprint{arXiv:astro-ph/0011147}.

\bibitem[{\citenamefont{Bowen et~al.}(2001)\citenamefont{Bowen, Hansen,
  Melchiorri, Silk, and Trotta}}]{Bowen:01}
\bibinfo{author}{\bibfnamefont{R.}~\bibnamefont{Bowen}},
  \bibinfo{author}{\bibfnamefont{S.}~\bibnamefont{Hansen}},
  \bibinfo{author}{\bibfnamefont{A.}~\bibnamefont{Melchiorri}},
  \bibinfo{author}{\bibfnamefont{J.}~\bibnamefont{Silk}}, \bibnamefont{and}
  \bibinfo{author}{\bibfnamefont{R.}~\bibnamefont{Trotta}},
  \bibinfo{journal}{submitted to MNRAS}  (\bibinfo{year}{2001}),
  \eprint{arXiv:astro-ph/0110636}.

\bibitem[{\citenamefont{Efstathiou}(2001)}]{efs:01}
\bibinfo{author}{\bibfnamefont{G.}~\bibnamefont{Efstathiou}}
  (\bibinfo{year}{2001}), \eprint{arXiv:astro-ph/0109151}.

\bibitem[{\citenamefont{Hu and Sugiyama}(1995)}]{HuandSugiyama}
\bibinfo{author}{\bibfnamefont{W.}~\bibnamefont{Hu}} \bibnamefont{and}
  \bibinfo{author}{\bibfnamefont{N.}~\bibnamefont{Sugiyama}},
  \bibinfo{journal}{Phys. Rev.} \textbf{\bibinfo{volume}{D51}},
  \bibinfo{pages}{2599} (\bibinfo{year}{1995}),
  \eprint{arXiv:astro-ph/9411008}.

\bibitem[{\citenamefont{Knox}(1995)}]{knox:95}
\bibinfo{author}{\bibfnamefont{L.}~\bibnamefont{Knox}}, \bibinfo{journal}{Phys.
  Rev. D} \textbf{\bibinfo{volume}{52}}, \bibinfo{pages}{4307}
  (\bibinfo{year}{1995}), \eprint{arXiv:astro-ph/9504054}.

\bibitem[{\citenamefont{Bond et~al.}(1999)\citenamefont{Bond, Efstathiou, and
  Tegmark}}]{EBT}
\bibinfo{author}{\bibfnamefont{J.~R.} \bibnamefont{Bond}},
  \bibinfo{author}{\bibfnamefont{G.}~\bibnamefont{Efstathiou}},
  \bibnamefont{and} \bibinfo{author}{\bibfnamefont{M.}~\bibnamefont{Tegmark}},
  \bibinfo{journal}{MNRAS} \textbf{\bibinfo{volume}{304}}, \bibinfo{pages}{75}
  (\bibinfo{year}{1999}), \eprint{arXiv:astro-ph/9702100}.

\bibitem[{\citenamefont{Press et~al.}(1992)}]{Numerical:92}
\bibinfo{author}{\bibfnamefont{W.~H.} \bibnamefont{Press}}
  \bibnamefont{et~al.}, \emph{\bibinfo{title}{Numerical Recipies in Fortran.
  The Art of Scientific Computing}} (\bibinfo{publisher}{Cambridge {UP}},
  \bibinfo{year}{1992}), \bibinfo{edition}{2nd} ed.

\end{thebibliography}

\end{document}